\documentclass[amsmath,amssymb,aps,pre,preprint,groupedaddress]{revtex4-2}

\usepackage{graphicx}
\usepackage{dcolumn}
\usepackage{bm}
\usepackage{ulem}
\usepackage{xcolor,soul}

\usepackage{longtable}

\newcommand*\Laplace{\mathop{}\!\mathbin\bigtriangleup}

\begin{document}

\title{Langevin and Navier-Stokes Simulation of Three-Dimensional Protoplasmic Streaming}



\author{Shuta Noro \textit{$^{1}$}}
\author{Satoshi Hongo \textit{$^{1}$}}
\author{Shinichiro Nagahiro \textit{$^{1}$}}\email[]{nagahiro@sendai-nct.ac.jp}
\author{Hisatoshi Ikai \textit{$^{1}$}}
\author{Hiroshi Koibuchi \textit{$^{2}$}}\email[]{koi-hiro@sendai-nct.ac.jp; koibuchi@gm.ibaraki-ct.ac.jp }
\author{Madoka Nakayama \textit{$^{3}$}}
\author{Tetsuya Uchimoto \textit{$^{4,5}$}}
\author{Gildas Diguet \textit{$^{6}$}}

\affiliation{
$^{1}$\quad~National Institute of Technology (KOSEN), Sendai College, 48 Nodayama, Medeshima-Shiote, Natori-shi, Miyagi 981-1239, Japan \\
$^{2}$\quad~National Institute of Technology (KOSEN), Ibaraki College, 866 Nakane, Hitachinaka, Ibaraki 312-8508, Japan \\
$^{3}$\quad~Research Center of Mathematics for Social Creativity, Research Institute for Electronic Science, Hokkaido University, Sapporo, Japan \\
$^{4}$\quad~Institute of Fluid Science (IFS), Tohoku University, 2-1-1 Katahira, Aoba-ku Sendai 980-8577, Japan \\
$^{5}$\quad~ELyTMaX, CNRS-Universite de Lyon-Tohoku University, 2-1-1 Katahira, Aoba-ku Sendai 980-8577, Japan \\
$^{6}$\quad~Micro System Integration Center, Tohoku University, 6-6-01 Aramaki-Aza-Aoba, Aoba-ku Sendai 980-8579, Japan%
}



\begin{abstract}
In this paper, we report the numerical results obtained using the Langevin Navier-Stokes (LNS) simulation of the velocity distribution of three-dimensional (3D) protoplasmic streaming in plant cells, such as those of {\it Nitella flexilis}. The LNS simulations are performed on 3D cylinders discretized by regular cubes in which fluid velocities are activated by boundary velocities parallel and nonparallel to the longitudinal direction and a random Brownian force with strength $D$. We find that, for a finite $D$, the velocity distribution $h(V), V\!=\!|\vec{V}|$, has two different peaks at a small non-zero $V$ and a finite $V$, and the distribution $h(V_z)$ for $|V_z|$ along the longitudinal direction also has a peak at finite $V_z$. These results are in good agreement with the reported velocity distributions observed using laser Doppler velocimetry. Moreover, we study the effects of the Brownian force on biological material mixing and find that mixing along the $\vec{V}$ direction enhanced by the nonparallel circular motion is further improved by the Brownian force  in the experimentally relevant region of $D$. In addition, the experimentally relevant $D$ is found to be consistent with the expectation from the fluctuation dissipation relation between  the random stress and viscosity in the LNS equation of Landau and Lifschitz for incompressible fluids.
\end{abstract}
%

\maketitle


\section{Introduction\label{intro}}
\begin{figure}[h!]
\begin{center}
\includegraphics[width=11.5cm]{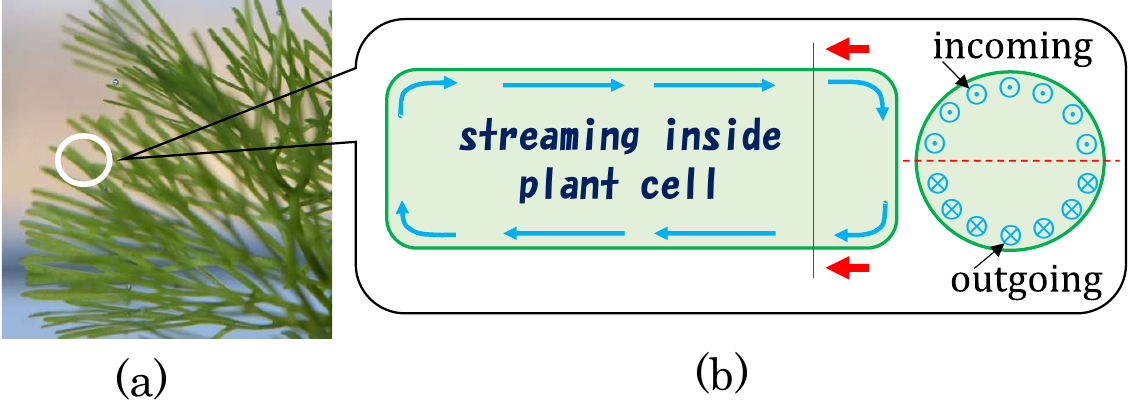}

\caption{ (a) Plant in water (image obtained using an optical camera), (b) illustration of the boundary flow that causes streaming inside cells, and the flow directions at the boundary region in a cell section. The symbol $\odot$ ($\otimes$) indicates the direction of velocity of the incoming side (outgoing to the opposite side).
	 \label{fig-1}
}
\end{center}
\end{figure}

The diameter of plant cells such as those of {\it Nitella flexilis} can be as high as 1 $({\rm mm})$ in water. Flow in these cells, termed protoplasmic streaming, has recently attracted considerable interest from researchers in biology and agriculture \cite{VLubics-Goldstein-Protoplasma2009,ShimmenYokota-COCB2004,TominagaIto-COPB2015}. Figures \ref{fig-1}(a) and (b) show an optical image of a plant in water and illustrate the streaming and its direction inside a cell, respectively.

The flow in plant cells is driven by a molecular motor in which a myosin molecule moves along actin filaments. Hence, the flow activation mechanism is the same as that in animal cells \cite{Squires-Quake-RMP2005,McintoshOstap-CSatAG2016,Astumian-Science2020,Julicher-etal-RMP1997}. Recently, Tominaga et al. reported that the size of a plant depends on the streaming velocity, which implies that the velocity of myosin molecules determines plant size \cite{TominagaIto-COPB2015}. To date, protoplasmic streaming has been extensively studied using experimental and theoretical techniques, including fluid dynamics simulations \cite{Goldstein-etal-PRL2008,Goldstein-etal-PNAS2008,Goldstein-etal-JFM2010,Raymond-Goldstein-IF2015,Kikuchi-Mochizuki-PlosOne2015,Niwayama-etal-PNAS2010}.

\begin{figure}[t]
\begin{center}
\includegraphics[width=10.5cm]{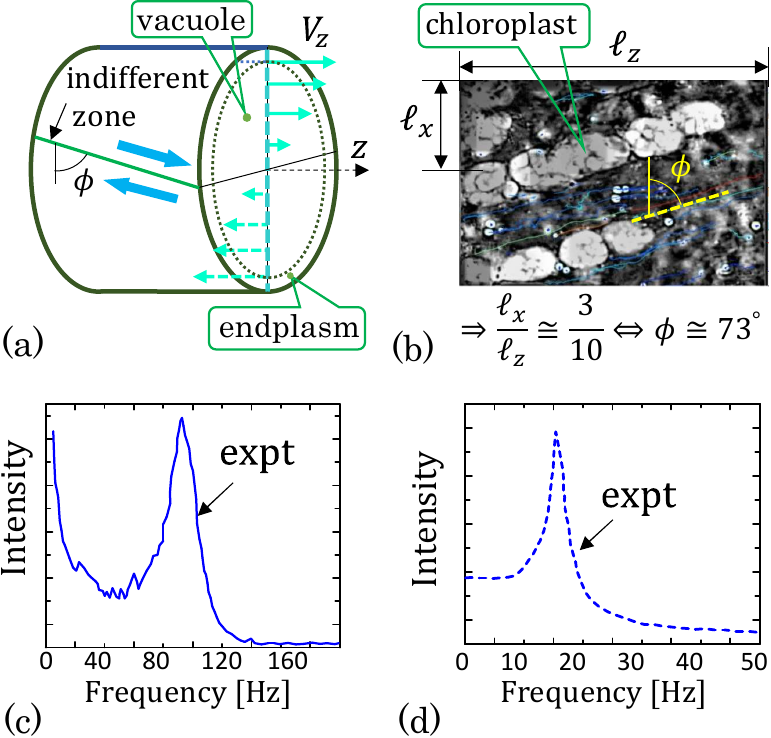}
\caption{ (a) Flow velocity ${\vec V}$  in vaculoe inside a cell, almost constant velocity in the endplasm \cite{Kamiya-Kuroda-1956,Kamiya-Kuroda-1958},  (b) photograph of {\it Nitella axilliformis Imahori} taken at the inner wall of an internodal cell \cite{Photo-Nittela}, (c) scattered light intensity vs. Doppler shift frequency obtained using a laser light scattering technique, where the peak at 93 Hz corresponds to a velocity of 72 $({\rm \mu m/s})$ \cite{Mustacich-Ware-BJ1976}, and (d) another reported scattered light intensity in \cite{Mustacich-Ware-BJ1977}. The Doppler-shift frequency depends on the scattering angle. The solid line in (a) denotes the indifferent zone, where two opposite boundary velocities are in contact with each other. The angle $\phi$ of the indifferent zone of {\it Nitella axilliformis Imahori} is estimated to be approximately $\phi\!=73^\circ$ from $\ell_x/\ell_z\!\simeq\!3/10$ in (b), where the $z$ direction is parallel to the longitudinal axis of the cell. }
 \label{fig-2}
\end{center}
\end{figure}

Kamiya and Kuroda first measured the flow velocity $V_z({\rm \mu m/s})$ along the longitudinal direction of a cell by using an optical microscope ($V_z$ in Fig. \ref{fig-2}(a)) \cite{Kamiya-Kuroda-1956,Kamiya-Kuroda-1958}. Corresponding physical quantities have also been reported; the kinematic viscosity is approximately 100 times larger than that of water \cite{Kamiya-Kuroda-1973,Kamiya-1986,Tazawa-pp1968}. In Ref. \cite{Pickard-CJB1971}, Pickard reported further observations of streaming including the angular and radial variations in the velocity with theoretical analyses. The flow direction along the side of the cell boundary is not always parallel to the longitudinal axis; rather, it is twisted, forming an indifferent zone (Fig. \ref{fig-2}(a)). Angle $\phi$ of the indifferent zone of {\it Nitella axilliformis Imahori} in Fig. \ref{fig-2}(b) is approximated to be $\phi\!=73^\circ$ \cite{Photo-Nittela}. 
Subsequently, using magnetic resonance velocimetry on cylinder cross sections, Goldstein et al. measured the velocity and reported the positional dependence of several different lines on the cross sections  with a theoretical study on the Stokes equation \cite{Goldstein-etal-JFM2010}. 
They also theoretically studied the flow field by combining the Stokes equation 
 and an advection-diffusion equation to describe the mixing of biological materials using a new variable concentration \cite{Goldstein-etal-PRL2008,Goldstein-etal-PNAS2008,Raymond-Goldstein-IF2015}.
Their results agreed well with the experimental results and they reported that biological material mixing is enhanced by the rotational boundary flow 
 \cite{Goldstein-etal-PRL2008,Goldstein-etal-PNAS2008,Goldstein-etal-JFM2010,Raymond-Goldstein-IF2015}.

Approximately 25 years after Kamiya and Kuroda's measurements, Mustacich and Ware observed streaming using a laser-light scattering technique \cite{Mustacich-Ware-PRL1974,Mustacich-Ware-BJ1976, Mustacich-Ware-BJ1977,Sattelle-Buchan-JCS1976}. They reported that the scattered light spectra exhibit two different peaks at $V\!\to\! 0$ and $V\!\not=\! 0$ (Fig. \ref{fig-2}(c))
and single peak at $V\!\not=\! 0$ (Fig. \ref{fig-2}(d)).
The peak velocity at $V\!\not=\! 0$ corresponds to the streaming velocity. 
It should be noted that the Brownian motion of biological materials influences the fluid velocities not only in the $V\!\to\!0$ limit but also for all the velocity ranges. This is the reason for the appearance of velocities in a range larger than the peak at finite velocity in the velocity distributions.

Recently, the peaks were numerically reproduced in Refs. \cite{Egorov-etal-POF2020,Noro-etal-2021} by simplifying the three-dimensional (3D) streaming to two-dimensional (2D) Couette flow and using the Langevin Navier-Stokes (LNS) equation. However, the 3D streaming nature, such as circulation at the cell boundary, was modified to be parallel to the longitudinal direction in the 2D simulations. Therefore, 3D simulations are preferable for a better understanding of the streaming.  

Here, we discuss the basic assumptions of the proposed model. Fluid flow inside the vacuole of a plant cell includes many biological materials, some of which can scatter laser light \cite{Mustacich-Ware-PRL1974,Mustacich-Ware-BJ1976, Mustacich-Ware-BJ1977,Sattelle-Buchan-JCS1976}. However, the sizes of biological materials are not uniform, and are expected to range from the size of molecules ($\simeq 10\, ({\rm n m})$) to the size of chloroplasts ($\simeq 5 \,({\rm \mu m})$) or larger, and the density of the mass is comparable to that of water. Therefore, we simply regard the streaming as a fluid flow described by the LNS equation rather than as a colloidal suspension, which is described by the equations of motion of particles, such as those in Brownian Dynamics \cite{Hossain-etal-POF2022}.

In this study, we attempt to simulate the experimentally observed effects of the Brownian motion of biological materials on the flow field using the LNS equation without biological materials. The primary objective is to numerically reproduce experimentally observed and reported velocity distributions.  The LNS equation in this paper is included in the framework of the LNS equation of Landau and Lifschitz implying that velocity field and pressure thermally or hydrodynamically  fluctuate and that only the mean values of many samples are meaningful as observable quantities \cite{Landau-Lifschitz-StatPhys}. Fluid velocities are directly activated by  Gaussian random Brownian forces in our LNS equation; hence, the implementation of the Brownian motion of biological materials is different from that in other simulation schemes.  The term "Brownian force" is usually used as a random force to activate small particles; however, we slightly extend it and use it for the streaming fluid itself. We also check whether the Brownian force is well defined for hydrodynamic interactions in the sense that the total momentum is conserved, even though it is not of the form of an internal force between fluids.

Colloidal suspensions in fluids are simulated using Stokesian Dynamics or Brownian Dynamics \cite{Ermark-McCammon-JCP1978,Brady-McCammon-ARFM1988,Roux-PhysA1992}. In this technique, hydrodynamic interactions between particles are implemented. Fluid particles have been introduced as Dissipation Particle Dynamics (DPD) to simulate fluids with thermal fluctuations \cite{Hoogerbrugge-Koleman-EPL1992,Espanol-Warren-EPL1995,Espanol-EPL1997}. Thermal fluid fluctuations have also been studied using the so-called Immersed Boundary Method for biological systems, in which external forces, including the Brownian force, act on both the fluid and the immersed boundary \cite{Peskin-ActaNum2002,Kramer-Majda-SiamJAM2004,Kramer-Peskin-Atzberger-CMAME2008}. Our LNS simulation technique differs from that of the Immersed Boundary Method.

In the case of the lattice Boltzmann method (LBM), the velocity distribution functions are simulated, and physical quantities such as the velocity and pressure are calculated using these functions \cite{Succi-LBM2001,Ladd-PRL1993,Bhadauria-etal-POF2021,Fu-etal-POF2022,Inamuro-etal-POF1997}. Therefore, the variables solved in the LBM are different from those in the NS equation, although the NS equation can be derived from LBM equations under certain conditions \cite{Inamuro-etal-POF1997}. Moreover, the velocity distribution functions in the LBM are defined for a lump of fluid particles. Thus, the LBM shares a particle simulation scheme.  A Langevin equation is used as a numerical technique in particle physics for functions on a lattice \cite{KGWilson-PRD1985,Ukawa-Fukugita-PRL1985,Hofler-Schwarzer-PRE2000}, so we consider that the Brownian motions can be combined with the NS equation for fluids \cite{Kopp-Yanovsky-POF2022}.  However, Brownian particles are not introduced; instead, the velocity and pressure of  fluids are considered to  fluctuate by Brownian forces in the LNS equation, as mentioned above. Therefore, our simulation scheme is slightly different from the LBM.

\section{Methods\label{methods}}
\subsection{Langevin Navier--Stokes equation and discrete equation\label{LNS-equation}}
The LNS equation is given by a set of coupled equations for the velocity $\vec{V}\!=\!(V_x,V_y,V_z)({\rm m/s})$ and pressure $p({\rm Pa})$:
\begin{eqnarray}
	\label{NS-eq-org}
	\begin{split}
		&\frac{\partial {\vec V}}{\partial t}=-\left ({\vec V}\cdot \nabla\right){\vec V}-{\rho}^{-1} {\it \nabla} p +\nu \Laplace {\vec V} + {\vec \eta},\\
		&\nabla\cdot {\vec V}=0,
	\end{split}
\end{eqnarray}
where $\rho({\rm kg/m^3})$ and $\nu({\rm m^2/s})$ denote fluid density and kinematic viscosity, respectively \cite{Egorov-etal-POF2020,Noro-etal-2021}. The final term $\vec{\eta}({\rm m/s^2})$ on the right-hand side of the first equation corresponds to the random Brownian force per unit mass. 
Because of this random force, the variables $\vec{V}$ and $p$ fluctuate rapidly at a small distance, and the mean value $\langle \vec{V}\rangle$ can be obtained as an observable physical quantity, as mentioned in the introduction. The meaning of the symbol $\langle * \rangle$ is introduced in the following subsection from a simulation perspective.

The velocity $\vec{V}$ and pressure $p$ variables are used in Eq. (\ref{NS-eq-org}), which differs from the LNS equation for the flow function $\psi$ and vorticity $\omega$ in Ref.~\cite{Egorov-etal-POF2020}, where the condition $\nabla\cdot {\vec V}\!=\!0$ is exactly satisfied for all $t$. In contrast, this divergenceless condition should manifest as a constraint  in the time evolution of Eq. (\ref{NS-eq-org}). The discrete-time step in Eq. (\ref{NS-eq-org}) violates $\nabla\cdot {\vec V}\!=\!0$. The original marker and cell (MAC) method is a simple technique to resolve this problem \cite{McKee-etal-CandF2007}. However, the $\nabla\cdot {\vec V}\!=\!0$ is not always satisfied even at the convergent solution satisfying ${\partial {\vec V}}/{\partial t}=0$. Hence, a well-known simplified MAC (SMAC) method is used in this study. 

To solve Eq. (\ref{NS-eq-org}), we impose the steady state condition
\begin{eqnarray}
	\label{SS-condition}
	\frac{\partial {\vec V}}{\partial t}=0. 
\end{eqnarray}
To obtain ${\vec V}$ satisfying this condition, we numerically solve the following discrete equation with time step ${\it \Delta} t$:
\begin{eqnarray}
	\label{NS-eq-time-step}
	{\vec V}(t+{\it \Delta}t)= \vec{V}(t) +{\it \Delta} t \left[ \left (-{\vec V}\cdot \nabla\right){\vec V}(t)-{\rho}^{-1} {\it \nabla} p(t+{\it \Delta} t) +\nu \Laplace {\vec V}(t)\right] +\sqrt{2D{\it \Delta} t} \,\vec{g},
\end{eqnarray}
where we use the same symbol $t$ for discrete time in this equation as for real time $t$ in the original LNS equation in Eq. (\ref{NS-eq-org}). This difference in time and detailed information of the SMAC method used to obtain the solution to Eq. (\ref{NS-eq-time-step}), under the conditions given in Eq. (\ref{SS-condition}) are presented in Appendix \ref{append-A}. The Brownian fluctuation process introduced by $\vec{\eta}(t)$ shares the energy dissipation process via the fluctuation dissipation relation, which is discussed in the following subsection. Therefore, the dissipation process can be described simply by $\vec{\eta}(t)$, whereas the corresponding energy input process, supplied by the molecular motors on the cell surface, is highly complex. Therefore, the LNS equation Eq. (\ref{NS-eq-org}) is used for the simulation. The symbol $D$ denotes the Brownian force strength, which is different from the diffusion coefficient.

\subsection{Lattices for simulations and boundary conditions\label{lattices}}
\begin{figure}[ht]
\begin{center}
\includegraphics[width=11.5cm]{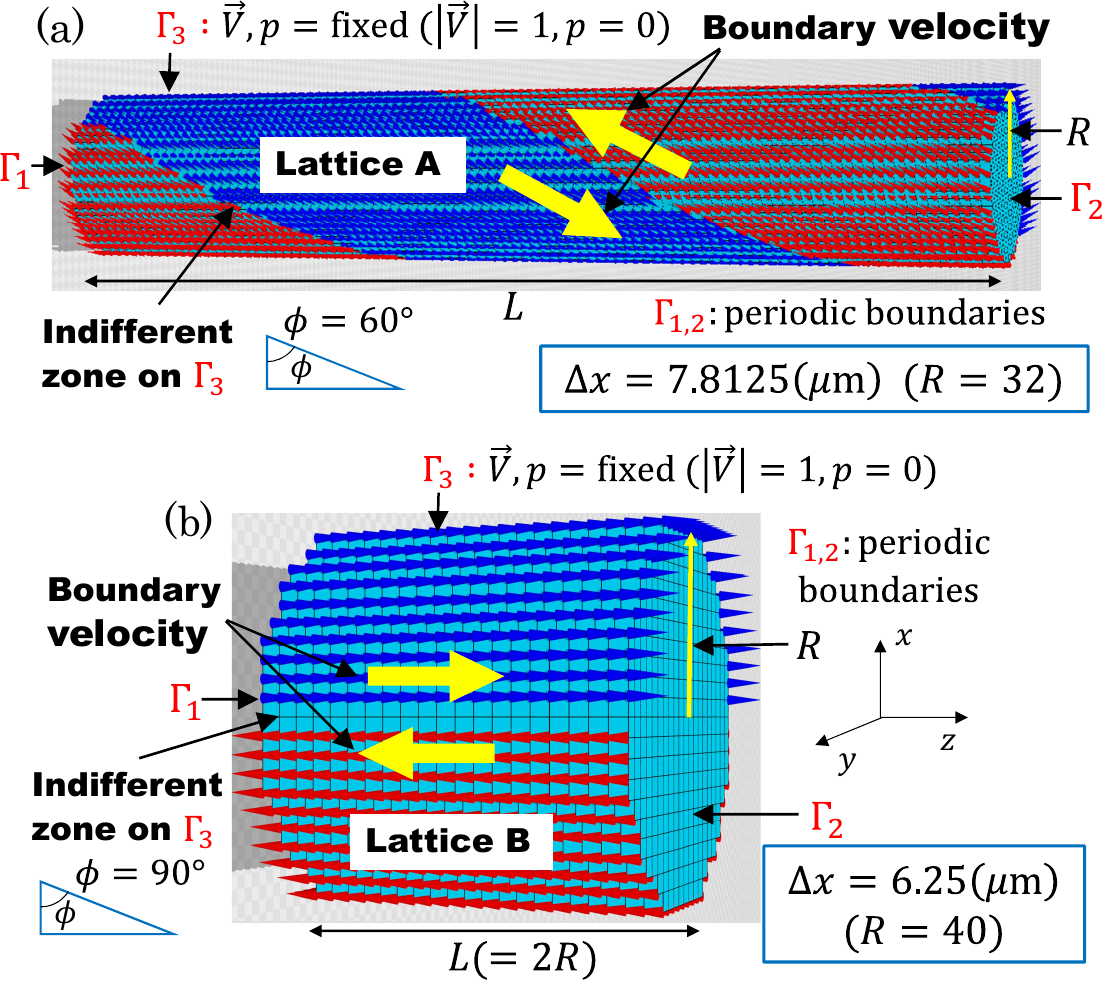}
\caption{ 3D cylindrical computational domains of (a) lattice A and (b) lattice B for streaming. The arrows in (a) and (b) indicate the directions of the boundary velocity that activates the streaming inside the cylinder, and the small cones represent the velocity directions. The length $L$ of the cylinder is determined such that the indifferent zone rotates once around boundary $\Gamma_3$: $L\!=\!2\pi(R+1)\tan\phi$ (with the unit of the lattice spacing ${\it \Delta}x$); consequently, the velocities and pressures on $\Gamma_1$ and $\Gamma_2$ are connected by the periodic boundary condition on lattice A. $L$ is fixed to $L\!=\!2R$ on lattice B.
	For clear visualization of the cones, the diameters of the cylinders in (a) $R\!=\!8$ and (b) $R\!=\!10$ are four times smaller than those (a) $R\!=\!32$ and (b) $R\!=\!40$ used for the simulations. The velocity $\vec{V}$ and pressure $p$ are fixed to $|\vec{V}|\!=\!1$ and $p\!=\!0$ in the simulation units on $\Gamma_3$ as boundary conditions for lattices A and B.
 \label{fig-3} }
\end{center}
\end{figure}
We show the details of the lattice construction for a cylindrical streaming domain in plant cells. The actual cell surface is soft and is expected to bend and fluctuate. However, it is relatively rigid compared to the surface of animal cells \cite{Passos-etal-POF2019}; therefore, we assume that the cylinder surface is rigid for simplicity. Thus, for the computational domain, we assume a 3D cylinder of radius $R$ and length $L$ (Fig. \ref{fig-3}). The lattice spacing ${\it \Delta}x$ is assumed to be ${\it \Delta}x\!=\!1$ in this and subsequent subsections, and $R{\it \Delta}x$ and $L{\it \Delta}x$ are written as $R$ and $L$, respectively, for simplicity. The regions indicated by the symbols $\Gamma_i (i\!=\!1,2,3)$ in the figure denote boundary surfaces.

The fluid is activated by the molecular motors on surface $\Gamma_3$, and the fluids are dragged along the boundary, as indicated by the two large arrows in Fig. \ref{fig-3}. The contact line along which two different velocities coexist is called the indifferent zone and divides $\Gamma_3$ into two domains. The angle of the zone is fixed to $\pi/3$ (or $60^\circ$) on lattice A, and the volume of the computational domain depends on this angle. An angle of $60^\circ$ is assumed, which is smaller than the actual angle in the plant cells, as shown in Fig. \ref{fig-2}(b), to save computational time. The length $L$ of the cylinder is fixed such that the indifferent zone rotates once around $\Gamma_3$. Therefore, the boundaries $\Gamma_1$ and $\Gamma_2$ are connected by a periodic boundary condition such that the velocities $\vec V$ and pressures $p$ on $\Gamma_1$ and $\Gamma_2$ are nearest neighbors to each other. The boundary velocity $\vec{V} (|\vec{V}|\!=\!1$ in the simulation units: Appendix \ref{append-B}) on $\Gamma_3$ on lattice A is fixed to be a unit tangential vector, and the orientation in one domain is opposite to that in the other, as shown in Fig. \ref{fig-3}(a). The tangential vectors are characterized by $|V_z|\!=\!\sin \phi (\phi\!=\!60^\circ)$, where $V_z$ is the $z$-component of $\vec{V}$. On lattice B, the boundary velocity is fixed to $|\vec{V}|\!=\!|V_z|\!=\!1$. Another boundary condition is $p\!=\!0$ for all points on $\Gamma_3$ of both lattices A and B. We simply fix $p\!=\!0$  for all points on $\Gamma_3$ of both lattices A and B. The boundary value of $p(t)$ is necessary because of $\nabla p(t)$ at the inner vertices for the calculation of a temporal velocity $\vec{V}^*(t)$ (Appendix \ref{append-A}).

\begin{figure}[h!]
\begin{center}
\includegraphics[width=9.5cm]{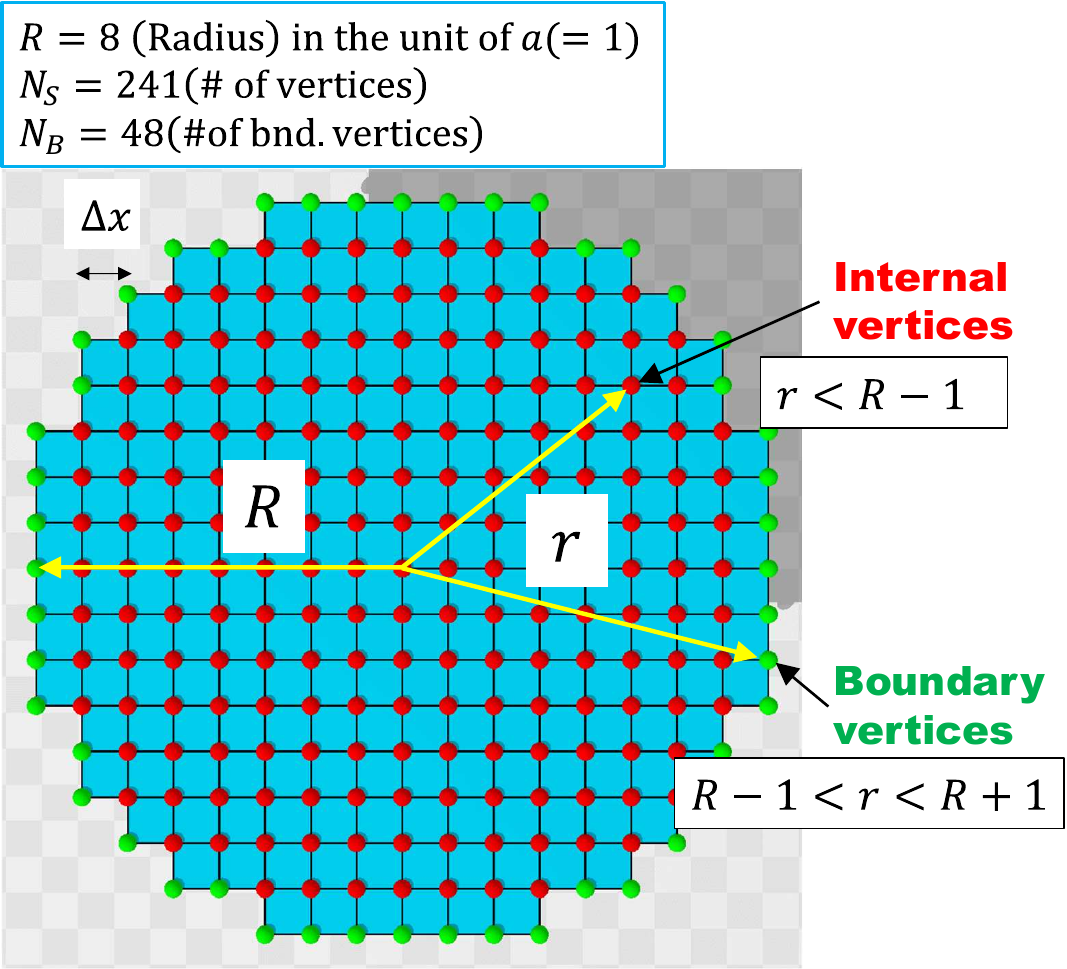}
\caption{The cross-section of a cylindrical lattice of size $R\!=\!8$, which is 4 to 5 times smaller than the $R\!=\!32$ and $R\!=\!40$ of the lattices for the simulations. The total number of vertices is $N_S\!=\!241$, which includes $N_B\!=\!48$ boundary vertices. The radius $r$ of the boundary vertices at which the velocity $\vec{V}$ and pressure $p$ are fixed is given by $R\!-\!1<r<R\!+\!1$, and the $r$ value of the internal vertices is given by $r<R\!-\!1$.
 \label{fig-4} }
\end{center}
\end{figure}
To explain the lattice structure, we show the cross-section of the cylinder (Fig. \ref{fig-4}(a)). The building block is a regular cube with lattice spacing ${\it \Delta}x$; therefore, the boundary shape is not circular. Let $r$ be the distance of a vertex from the center of the cross-section. Vertices in region $R\!-\!1\!<\!r\!<\!R\!+\!1$ form the boundary, whereas those in region $r\!<\!R\!-\!1$ are the internal points, where $R$ is the radius of the horizontal and vertical lines passing through the center of the cross-section. The geometries of lattices A and B are summarized in Table \ref{table-1}.

\begin{table}[b]
\caption{Two different lattice geometries for the simulations. The ratio $L/R$ is approximately $L/R\!\simeq\!11.2$ on lattice A, while it is exactly $L/R\!=\!2$ on lattice B, where $L$ is the cylinder length and $R$ is the radius with the unit of lattice spacing  ${\it \Delta}x(=\!1)$ (Fig. \ref{fig-3}(a) and (b)). The reason for the difference in $L/R$ between lattices A and B is discussed in the caption of Fig. \ref{fig-3}.
 }
\label{table-1}
\begin{center}
 \begin{tabular}{ccccccccccc }
 \hline
\vspace{-3mm}
Lattice & $\phi$ & $R$ && $L$ && Internal & & Boundary \\
  & &  && && vertices & & vertices \\
 \hline
A & $60^\circ$ & 32 && 359 && 1,153,800 && 66,240 \\
B & $90^\circ$ & 40 && 80 && 406,053 && 18,468 \\ 
 \hline
\end{tabular}
\end{center}
\end{table}

The size of both lattices is relatively small owing to the stochastic nature of the model, because many convergent configurations are necessary to obtain the mean values of the physical quantities:
\begin{eqnarray}
	\label{mean-of-Q}
	\langle Q\rangle=(1/ n_{\rm s})\sum_{i=1}^{n_{\rm s}}Q_i,
\end{eqnarray}
where $Q_i$ denotes the $i$th convergent configuration corresponding to the $i$th Gaussian random force $\vec{\eta}(t)$. The symbol $n_{\rm s}$ is the total number of convergent configurations, and $n_{\rm s}=1000$ for the calculation of $\vec{V}(r,\theta)$, which is shown in the following subsection, for all $D$ except $D\!=\!0$. For $D\!=\!0$, $n_{\rm s}$ should be $n_{\rm s}\!=\!1$ because no Gaussian random number is assumed in this case. For simplicity, brackets $\langle \cdot \rangle$ are not used for the mean values in this paper. Here, we emphasize that the numerical technique for obtaining physical quantities in this study and in Refs. \cite{Egorov-etal-POF2020,Noro-etal-2021} is based on Eq. (\ref{mean-of-Q}). This calculation technique is used to obtain the canonical ensemble averages of physical quantities in statistical mechanical simulations such as the Metropolis Monte Carlo simulation technique \cite{Metropolis-JCP-1953,Landau-PRB1976}. The aim of our LNS equation is not to determine the time evolution of the fluid flow on a long time scale, but simply to obtain the equilibrium velocity configuration under Brownian impulses that play a role in thermal fluctuations.

\subsection{ Velocity distributions $h(V)$, $h(V_z)$ and the radial dependence of $V_z$ \label{calculation}}
\begin{figure}[ht]
\begin{center}
\includegraphics[width=8.5cm]{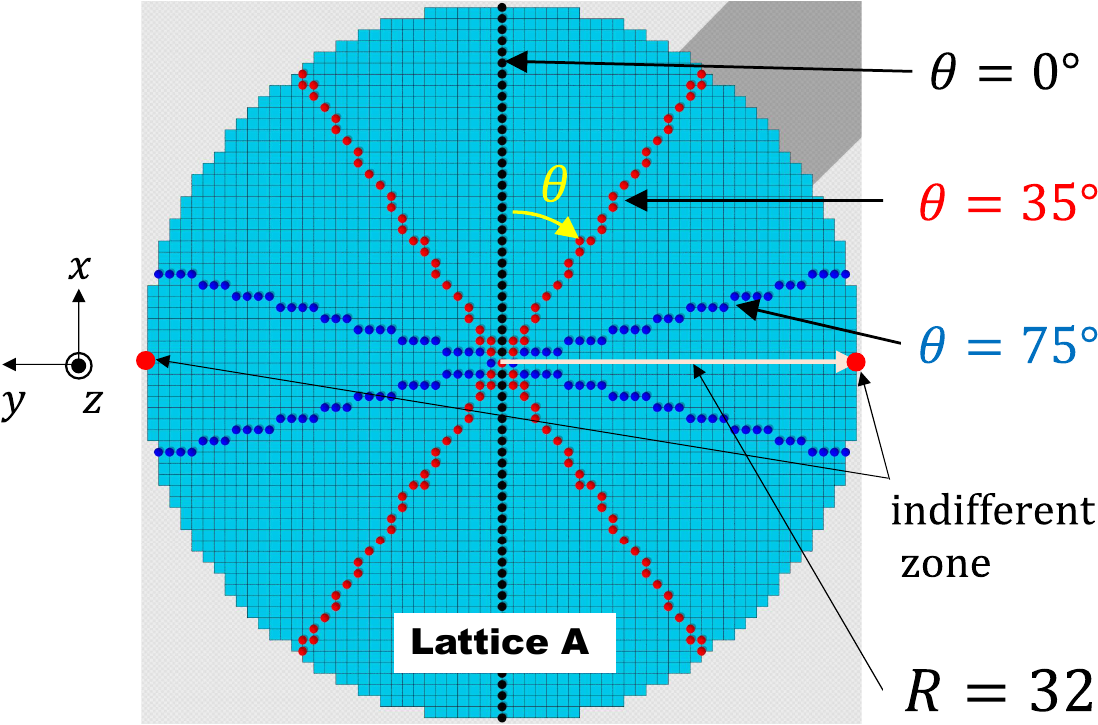}
\caption{Lattice sites along the lines of three different angles $\theta$ for calculating $V_z(r,\theta)$. The line at $\theta\!=\!0^\circ$ is a vertical line along the $x$ axis, while those at $\theta\!=\!35^\circ$ and $\theta\!=\!75^\circ$ are two symmetric lines due to the reflection symmetry $\theta\to-\theta$. The lattice section corresponds to lattice A, whose radius is given by $R\!=\!32$ with the unit of lattice spacing ${\it \Delta}x(=\!1)$.
 \label{fig-5} }
\end{center}
\end{figure}
The dependence of $V_z(r,\theta)$ on $r$ is numerically measured along the lines in Fig. \ref{fig-5} with angles $\theta\!=\!0^\circ,35^\circ$, and $75^\circ$ on the cross-section at $z\!=\!L/2$ at the midpoint of the cylinder. These angles are almost the same as those assumed in Ref. \cite{Goldstein-etal-JFM2010}.
Only a single cross-section is used for the numerical measurements of $V_z(r,\theta)$ on lattice A because the boundary velocities are rotating and there is no equivalent cross-section along the longitudinal direction of the cylinder, although the cross-sections at $z\!=\!0$ and $z\!=\!L$ are almost equivalent owing to the periodicity of the circulation (Fig. \ref{fig-3}(a)). In the case of lattice B, all the cross-sections are equivalent; however, we also use the cross-section at $z\!=\!L/2$ to numerically measure $V_z(r,\theta)$, as in the case of lattice A.

The experimentally observed laser light scattering intensity in Fig. \ref{fig-2}(c) is considered to correspond to the velocity distribution $h(V_z)$, which has no $\theta$ dependence. Therefore, we calculate $h(V_z)$ and $h(V)$ for all cross sections. Detailed information regarding the $h(V)$ and $h(V_z)$ calculation techniques is provided in Appendix \ref{append-C}.

\subsection{Input parameters}
The physical parameters that characterize protoplasmic streaming are the density $\rho_e ({\rm kg/m^3})$, kinematic viscosity $\nu_e ({\rm m^2/s})$, boundary velocity $V_e ({\rm m/s})$, and diameter of the cell $d_e ({\rm m})$ (Table \ref{table-2}). The symbols with subscript $e$  except for $R_e$ denote experimentally observed or observable quantities.
\begin{table}[htb]
\caption{Physical parameters $\nu_{e}, V_{e},$ and $d_{e}$ corresponding to the protoplasmic streaming in plant cells, expressed in physical units. The estimated Reynolds number is $R_e\!=\!V_ed_e/\nu_e\!=\!2.5\times 10^{-4}$. The symbols with subscript $e$  except for $R_e$ denote experimentally observed or observable quantities.
\label{table-2}} 
\begin{center}
 \begin{tabular}{ccccccccccccccc}
 \hline
 \hline
 $\rho_e ({\rm kg/m^3})$ && $\nu_{e}~({\rm m^2}/{\rm s})$ && $V_{e}~({\rm \mu m}/{\rm s})$ && $d_{e}~({\rm \mu m})$ \\
 \hline
 $1\times 10^{3}$  && $1\times 10^{-4}$  && $50$ && $500$   \\
\hline
\hline
 ${\it \Delta}x ({\rm A})~({\rm \mu m})$ && ${\it \Delta}x ({\rm B})~({\rm \mu m})$ && ${\it \Delta}t~({\rm s})$ \\
\hline
 $7.8125$ && $6.25$ && $2\!\times\!10^{-8}$  \\
\hline
\hline
\end{tabular}
\end{center}
\end{table}
These values are provided in Refs. \cite{Kamiya-Kuroda-1956,Kamiya-Kuroda-1958,Kamiya-Kuroda-1973,Kamiya-1986,Tazawa-pp1968} and are the same as those assumed in the 2D LNS simulations in Refs. \cite{Egorov-etal-POF2020,Noro-etal-2021}. The lattice spacings ${\it \Delta}x ({\rm A})$ and ${\it \Delta}x ({\rm B})$ of lattices A and B and the time step ${\it \Delta}t$ are listed in Table \ref{table-2}. Note that the boundary velocity $V_{e}$ is extremely small, implying that the first term or advection term on the right-hand side of Eq. (\ref{NS-eq-org}) is negligible. Therefore, the LNS equation in Eq. (\ref{NS-eq-org}) or (\ref{NS-eq-time-step}) can be called the Langevin–Stokes equation, and the fact that the advection term is negligible implies that the analyses reported in Refs. \cite{Goldstein-etal-PRL2008,Goldstein-etal-PNAS2008} are suitable for protoplasmic streaming. The Reynolds number of the streaming can be estimated such that $R_e\!=\!V_ed_e/\nu_e\!=\!2.5\times 10^{-4}$.

Using the factors $\alpha, \beta$ and $\lambda$ for the unit change (see Appendix \ref{append-B}), we obtain the parameters in Table \ref{table-3} in the simulation units, which are used in the simulations in this study. 
\begin{table}[htb]
\caption{Parameters assumed in the simulations; these values are given in the simulation units. 	 \label{table-3} }
\begin{center}
 \begin{tabular}{cccccccccccc }
 \hline
 Lattice && $\rho_0~[\frac{\rm \lambda kg}{\rm (\alpha m)^3}]$ && $\nu_0~[\frac{\rm (\alpha m)^2}{\rm \beta s}]$ &&\ $V_0~[\frac{\rm \alpha m}{\rm \beta s}]$ && ${\it \Delta}x_0~[{\rm \alpha m}]$ && ${\it \Delta}t_0~[{\rm \beta s}]$ \\
 \hline
 A && $1\times 10^{-3}$  && $1\times 10^6$  && 1 && 3.90625 && $5\times10^{-7}$ \\
B && $1\times 10^{-3}$  && $1\times 10^6$  && 1 && 3.125 && $5\times10^{-7}$ \\
 \hline
\end{tabular}
\end{center}
\end{table}
Note that $D$ is expressed using the simulation units $({\rm (\alpha m)^2/(\beta s)^3})$ in this study (Appendix \ref{append-B}). The physical scales corresponding to $D$ are discussed in the following subsections.  

We consider $D$ as the only parameter that can be varied in the simulations. The target phenomenon is characterized by $(\rho_e, \nu_{e}, V_{e}, d_{e})$ in Table \ref{table-2}, and the corresponding parameters $(\rho_0, \nu_{0}, V_{0}, {\it \Delta} x_{0})$ in the simulation units are listed in Table \ref{table-3}. Thus, if we fix ${\it \Delta} t_{0}$ to a certain value, then the variable parameter is only $D$, and we should determine a $D$ suitable for the thermal fluctuations of the target system by comparing the simulated $h(V_z)$ with the experimentally observed and reported $h(V_z)$.

\subsection{Fluctuation and dissipation relation and eligibility of the LNS simulation \label{FD-ReL}}
It is well known that myosin molecules activate chloroplasts at the cell boundary, and chloroplasts drive the streaming, including that of biological materials inside the cells (Fig. \ref{fig-2}(b)). From a microscopic perspective, biological materials collide with water molecules and lose kinetic energy.  This collision process for energy dissipation is the same as that in (or can also be viewed as) the Brownian motion of biological materials owing to the fluctuation-dissipation relation, which will be discussed in this subsection.

The LNS equation was introduced by Landau and Lifschitz in the context of the fluctuation-dissipation relation \cite{Landau-Lifschitz-StatPhys}. Landau and Lifschitz assumed fluctuations in all fluid mechanical quantities such as velocity, density, and pressure. Fluctuations in these quantities are introduced via divergence $\partial s_{ab}/\partial x_b(=\!\sum_{b=1}^3\partial s_{ab}/\partial x_b)$ of a random stress $s_{ab}$, which is a component of the stress tensor $\sigma^\prime_{ab}$ and plays a role as a source of the fluctuations. In this sense, the $a$-th component of the  $\vec{\eta}(\vec{r},t)$  term  in Eq. (\ref{NS-eq-org})  corresponds to $\partial s_{ab}/\partial x_b$ such that
\begin{eqnarray}
	\label{LL-corespondence}
	\eta_i^a(t) \leftrightarrow \rho^{-1}\partial s_{ab}/\partial x_b,
\end{eqnarray}
where fluctuations are assumed only in the velocity and pressure because of the assumption of incompressibility. The right-hand side describes a force of interaction between fluids, whereas the left-hand side describes the activation force of fluids at lattice site $i$ corresponding to $\vec{r}$, and it appears to be an external force that can violate the momentum conservation in the flow field. However, $\vec{\eta_i}$ is given by a Gaussian random number with mean zero (Appendix \ref{append-A}). Therefore, $\langle\sum_i \vec{\eta_i} \rangle\left(=\!\sum_i \langle\vec{\eta_i} \rangle\right)\!=\!\vec{0}$ is satisfied in the sense of mean value. Thus, the correspondence in Eq. (\ref{LL-corespondence}) is well defined, which is numerically confirmed  in a later subsection. 

We assume the relation
\begin{eqnarray}
\label{Gaussian-random-force}
\langle \eta_i^a(t)\eta_j^b(t^\prime)\rangle = 2D\delta_{ij}\delta^{a b}\delta(t-t^\prime),
\end{eqnarray}
where $i$ and $j$ denote space points and $a$ and $b$ represent the directions. 	The symbol $D$ is not the diffusion coefficient but the strength of the Brownian force. No information on viscosity is included on the right-hand side, which describes only the properties of Brownian force. However, we call this relation the fluctuation dissipation relation, because the Brownian force strength $D$ in this relation is related to the viscosity via the above-mentioned  correspondence in Eq. (\ref{LL-corespondence}) and the fluctuation dissipation relation of the stress tensor $s_{ab}$ of Landau and Lifschitz, as described in Appendix \ref{append-D}. This relation  in Eq. (\ref{Gaussian-random-force}) is identical to that introduced in Ref. \cite{Egorov-etal-POF2020}, except for the space dimension, and implies that the hydrodynamic fluctuation activation force is negligibly small in its variation in the space and time directions compared with the characteristic scales in space and time in the reference physical system. In our case, the physical system is the fluid velocity $\vec{V}$ of protoplasmic streaming activated by the boundary flow, and the randomly changing forces originate from the thermally fluctuating water molecules, which activate biological materials flowing in the streaming with the same velocity  $\vec{V}$. Thus, it is reasonable to study the fluctuations in $\vec{V}$ using the LNS equation instead of particle dynamics for the Brownian motion of biological materials. Temperature $T$ is not explicitly included in the LNS equation; however, the random force strength $D$ is proportional to $k_BT$ under the correspondence in Eq. (\ref{LL-corespondence}) (see Appendix \ref{append-D}). Therefore, thermal fluctuations increase as $D$ increases.

Note that  $\delta_{ij}$ is used in Eq. (\ref{Gaussian-random-force}) instead of $\delta(\vec{r}\!-\!\vec{r}^{\,\prime})$ for the spatial correlation of $\vec{\eta}(\vec{r},t)$ in the fluctuation dissipation relation of $s_{ab}$ in Ref. \cite{Landau-Lifschitz-StatPhys}  (Appendix \ref{append-D}). In Ref. \cite{Landau-Lifschitz-StatPhys}, $\delta(\vec{r}\!-\!\vec{r}^{\,\prime})$ is identified with $\delta_{ij}/{\rm v}_i$ in the limit of zero volume ${\rm v}_i\!\to\!0$. This replacement  is suitable for the numerical simulations of the LNS equation in Eq. (\ref{NS-eq-org}). In particular, the Brownian force strength $D$ in Eq. (\ref{Gaussian-random-force})  depends on  ${\it \Delta}x$  and ${\it \Delta}t$; hence, if we change these parameters, $D$ should also be changed to obtain the same results \cite{Egorov-etal-POF2020,Noro-etal-2021}. This ${\it \Delta}x$-dependence of $D$ originates from the notation $D$ in Eq. (\ref{Gaussian-random-force}), which corresponds to $\bar D/({\it \Delta}x)^d (d\!=\!3)$ with the strength $\bar D$ regarding the spatial correlation   $\delta(\vec{r}\!-\!\vec{r}^{\,\prime})$.  Therefore, we first fix the lattice spacing ${\it \Delta}x$ to be sufficiently small for numerical accuracy depending on the lattice size, and then the discrete time step ${\it \Delta}t$ is fixed sufficiently small for a combination with sufficiently large $D$, maintaining a fast convergence,  and varying $D$ in the simulations to find a suitable $D$ for the streaming under consideration.  The experimentally relevant value of $D$ observed in the simulations can be compared with the expectation from the correspondence between the relation in Eq. (\ref{Gaussian-random-force}) and the fluctuation dissipation relation of $s_{ab}$ for the LNS equation of Landau and Lifschitz, we confirm that $D$ suitably assumed in the simulations in this study is consistent with the expectation (Appendix \ref{append-D}).  

The discrete expression $\sqrt{2D{\it \Delta} t} \,\vec{g}$ for the random force in Eq. (\ref{NS-eq-time-step}) originates from the stochastic nature of $\vec{\eta}(\vec{r},t)$ characterized by Eq. (\ref{Gaussian-random-force}). A detailed explanation of this expression $\sqrt{2D{\it \Delta} t} \,\vec{g}(t)$ is discussed from the viewpoint of impulse action (Appendix \ref{append-D}).

 Using the notion of this impulse action, we calculate the Brownian force acting on the fluid volume $({\it \Delta} x)^3$ to estimate the physical scales corresponding to $D$. The magnitude of the impulsive force, which changes the velocity $\vec{V}_{i}$ of the volume $({\it \Delta} x)^3$ at lattice site $i$,  is given by $\left\|\vec{\eta}_{i}(t)\right\|\!=\!\sqrt{6D/{\it \Delta} t}$, which is evaluated to $\left\|\vec{\eta}_{i}(t)\right\|^2\!=\!|\eta^1_{i}(t)|^2\!+\!|\eta^2_{i}(t)|^2\!+\!|\eta^3_{i}(t)|^2\!=\!3|\eta^a_{i}(t)|^2$ and $|\eta^a_{i}(t)|\!=\!\sqrt{2D/{\it \Delta} t}, (a\!=\!1,2,3)$ in terms of the squared mean value (Appendix \ref{append-D}). Note that the Brownian force $\vec{\eta}_{i}(t)$ at $i$ for an infinitesimal timescale is replaced by a finite constant  $\left\|\vec{\eta}_{i}(t)\right\|\!=\!\sqrt{6D/{\it \Delta} t}$ during the finite time ${\it \Delta} t$ under the same momentum transfer. In this sense, $\vec{\eta}_{i}(t)$ is not a classical force because impulse $\vec{H}_{i}(t; {\it \Delta} t)\!=\!\int_t^{t+{\it \Delta} t}\vec{ \eta}_{i}(t)dt\!=\!\vec{ \eta}_{i}(t){\it \Delta} t\!=\!\sqrt{2D{\it \Delta} t}\vec g_{i}(t)$ satisfies the relations in Eq. (\ref{Brownian-motion-dt}), which is typical of stochastic variables. Here, we replace $\vec{ \eta}_{i}(t){\it \Delta} t$ with classical force $\left\|\vec{\eta}_{i}(t)\right\|\!=\!\sqrt{6D/{\it \Delta} t}$ to estimate the physical scale corresponding to $D$. This replacement is equivalent to replacing the Gaussian random number $g_i^a(t)$ with a squared mean value of one in Eq. (\ref{NS-eq-time-step}).   Using the density of fluids $\rho$ and lattice spacing ${\it \Delta} x$,  we obtain the magnitude of the $D$-dependent Brownian force.
 \begin{eqnarray}
\label{BR-force}
 f_{\rm BR}(D)=\rho ({\it \Delta} x)^3\left\|\vec{\eta}_{i}(t)\right\|\!=\!\rho ({\it \Delta} x)^3\sqrt{6D/{\it \Delta} t}.
 \end{eqnarray}
For $D\!=\!100$, we estimate  $ f_{\rm BR}(D\!=\!100)$ as $f_{\rm BR}(100)\!=\!\rho_e ({\it \Delta} x)^3\left\|\vec{\eta}_{i}(t)\right\|\!=\!\rho_e ({\it \Delta} x)^3\sqrt{2D/{\it \Delta} t}\!\simeq\!2.1\!\times\! 10^{-11} ({\rm kg m/s^2})\!\simeq\!21({\rm p N})$ using $\rho_e\!=\!1\!\times\!10^3 ({\rm kg/m^3})$, ${\it \Delta} x\!=\!7.8125\!\times\!10^{-6}({\rm m})$ and ${\it \Delta} t\!=\!2\!\times\! 10^{-8}({\rm s})$ in Table \ref{table-2}, and $D\!=\!100({\rm \alpha^2m^2/\beta^3s^3})\!=\!6.25\!\times\! 10^{-6}({\rm m^2/s^3})$. This impulse Brownian force $f_{\rm BR}(100)\!\simeq\!21({\rm p N})$ remains unchanged if ${\it \Delta} t$ is replaced by ${\it \Delta}\tau$ in Eqs. (\ref{Integral-GRF}) and (\ref{Brownian-motion}) because of the relation in Eq. (\ref{D-correspondence}). Note also that  ${\it \Delta}V$ is independent of the lattice spacing ${\it \Delta} x$ even though  $f_{\rm BR}(D)$ depends on it.  This  $f_{\rm BR}$  changes velocity $\vec{V}_i(t)$ to $\vec{V}_i(t\!+\!{\it \Delta} t)$ by ${\it \Delta} V\!=\!\|\vec{V}_i(t\!+\!{\it \Delta} t)\|\!-\!\|\vec{V}_i(t)\|$ such that ${\it \Delta} V\!=\! f_{\rm BR}\rho_e^{-1} ({\it \Delta} x)^{-3}{\it \Delta} t\!=\!\left\|\vec{\eta}_{i}(t)\right\|{\it \Delta} t\!=\!\sqrt{6D{\it \Delta}t}\!\simeq\! 0.9 ({\rm \mu m/s})$. 

This ${\it \Delta}V$ is obtained by neglecting all other forces during ${\it \Delta} t\!=\!2\!\times\! 10^{-8} ({\rm s})$ and replacing $g_i^a(t)$ with $g_i^a(t)\!\to\!1$ as described above.  In general, the equilibrium configuration of $\vec{V}_i$ is influenced by the Brownian random force, and the velocity change can be evaluated even without the assumption that $g_i^a(t)\!\to\!1$. 
To evaluate $\vec{V}_i$ under an impulse $\vec{H}_{i}(t; {\it \Delta} t)\!=\!\vec{ \eta}_{i}(t){\it \Delta} t$, we consider the simplified equation
  \begin{eqnarray}
	\label{LG-eq-simplify}
	\frac{\partial {\vec V}_i}{\partial t}(t)={\vec \eta}_i(t)
\end{eqnarray}
obtained from Eq. (\ref{NS-eq-org}) by neglecting all the other terms. Using this expression, we have
\begin{eqnarray}
	\label{mean-product}
	\left\langle \frac{\partial {V}^a_i}{\partial t}(t)\frac{\partial {V}^a_i}{\partial t^\prime}(t^\prime)\right\rangle=\left\langle {\eta}^a_i(t) {\eta}^a_i(t^\prime)\right\rangle,\; (a\!=\!1,2,3).
\end{eqnarray}
Hence, $\left\langle \int_t^{t+{\it \Delta}t}\frac{\partial {V}^a_i}{\partial t}(t)dt\int_{t^\prime}^{t^\prime+{\it \Delta}t}\frac{\partial {V}^a_i}{\partial t^\prime}(t^\prime)dt^\prime\right\rangle\!=\!\left\langle\int_t^{t+{\it \Delta}t}\eta^a_i(t)dt\int_{t^\prime}^{t^\prime+{\it \Delta}t}\eta^a_i(t^\prime )dt^\prime\right\rangle$ because the expectation and integration are commutative. It is easy to determine that $\left\langle({\it \Delta} V_i^a)^2\right\rangle\!=\!2D{\it \Delta}t$   $\left(\Leftrightarrow\!\langle{\it \Delta} V_i^a\rangle\!=\!\sqrt{2D{\it \Delta}t}\right)$ from Eq. (\ref{Gaussian-random-force}) in the limit of $t^\prime\!\to\!t$, where ${\it \Delta} V_i^a\!=\!V_i^a(t\!+\!{\it \Delta}t)\!-\!V_i^a(t)$.  Thus, we have $\left\langle{\it \Delta} V_i\right\rangle\!=\!\sqrt{3}\left\langle{\it \Delta} V_i^a\right\rangle\!=\!\sqrt{6D{\it \Delta}t}$, which is identical to the classical estimate ${\it \Delta} V$.

The problem is how to observe the effects of the ${\it \Delta}V \!\simeq\!0.9 ({\rm \mu m/s})$, which is relatively small compared with the boundary velocity $V_e\!=\!50 ({\rm \mu m/s})$ in Table \ref{table-2}. However, ${\it \Delta}V \!\simeq\!0.9 ({\rm \mu m/s})$ is due to a single impulse during ${\it \Delta} t$, and ${\it \Delta}V$ increases proportionally to the total number of hits because the same Brownian impulse ($\Leftrightarrow$ the same $\vec{g}_i(t)$) is used at each ${\Delta} t$ until the convergent configuration is obtained, and hence, a non-trivial influence of the random force on $\vec{V}_i$ is expected. One possible solution is to determine its influence on the velocity distribution, which is the mean value of many samples and is independent of the $\pm$ direction. Thus, we consider that the estimate ${\it \Delta}V \!\simeq\!0.9 ({\rm \mu m/s})$ during ${\it \Delta} t$ is physically meaningful if the simulation results of the velocity distributions are consistent with the experimentally observed velocity distribution. Generally, if we find a suitable $D$ such that the corresponding simulated distribution of velocities is  consistent with the experimental data, we conclude that the LNS simulation technique under Eq. (\ref{Gaussian-random-force}), is suitable for studying protoplasmic streaming.   This is discussed in the following Section.

\section{Numerical results \label{numerical-results}}
\subsection{Velocity distribution for $D\!=\!0$ \label{D_zero}}
\begin{figure}[ht]
\begin{center}
\includegraphics[width=11.5cm]{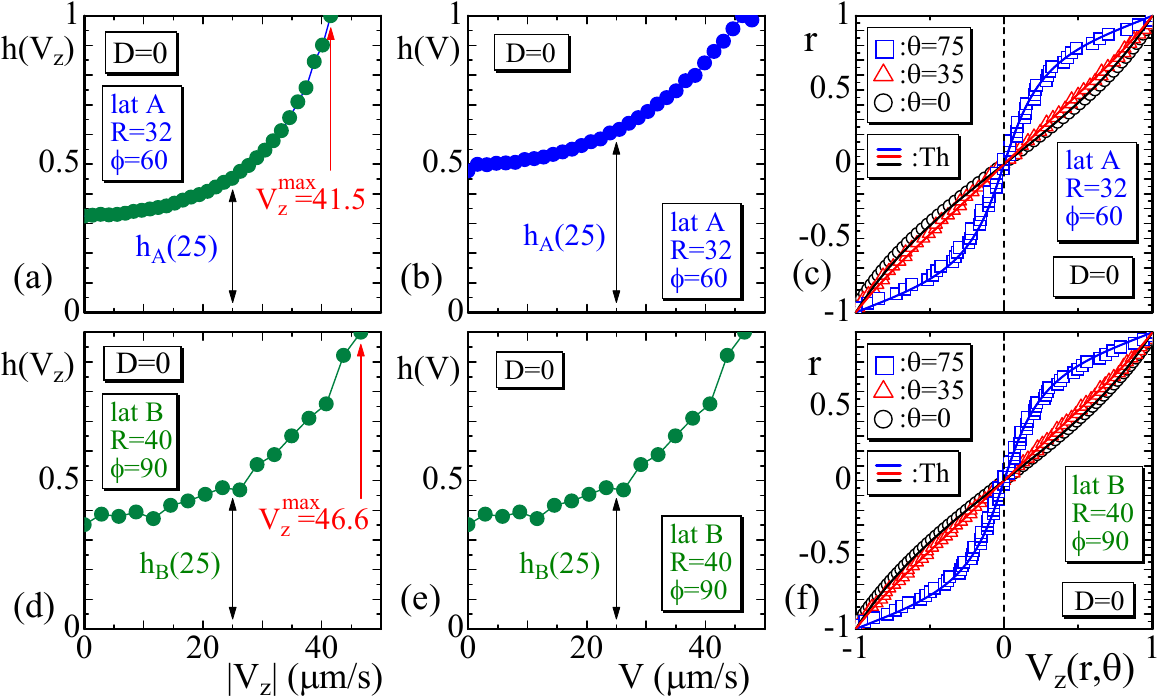}
\caption{ Velocity distributions (a) $h(V_z)$ vs. $|V_z|$, (b) $h(V)$ vs. $V$, (c) radial dependence of normalized $V_z(r,\theta)$ obtained on lattice A, and  (d), (e), (f) those obtained on lattice B at $D\!=\!0$. The results ($\bigcirc$) in (c) and (f) at $\theta\!=\!0$ correspond to the velocity distribution illustrated in Fig. \ref{fig-2}(a). The solid lines denoted as Th in (c) and (f) are drawn using Eq. (\ref{Pickard}). 
 \label{fig-6} }
\end{center}
\end{figure}
First, we introduce the results obtained for $D\!=\!0$ on lattice A. The distributions or histograms $h(V_z)$ of $|V_z|$ and $h(V)$ of $V$ are plotted in Fig. \ref{fig-6}(a) and (b), where the calculation technique for $h(V_z)$ and $h(V)$ is described in Appendix \ref{append-C}. The plotted data are calculated from a single convergent configuration of $\vec{V}$. Therefore, a slightly non-smooth behavior is observed in the data. The shape of $h(V_z)$ versus $|V_z|$ is nearly the same as that of $h(V)$ versus $V$. No peaks are observed in $h(V_z)$ and $h(V)$, which are determined only by the boundary fluid flow.
Figure \ref{fig-6}(c) shows the dependence of $V_z(r,\theta)$ on the distance $r$ from the center of the cross section  and $\theta$ (Fig. \ref{fig-5}). The solid lines show theoretical predictions reported in Ref. \cite{Pickard-CJB1971} given by
\begin{eqnarray}
	\label{Pickard}
	V_z(r,\theta)=\frac{2}{\pi}V\tan^{-1}\left[\frac{2(r/r_{\rm max})\sin (90^\circ-\theta)}{1-(r/r_{\rm max})^2}\right], \quad (0\leq r \leq r_{\rm max}) 
\end{eqnarray}
under the conditions $r_{\rm max}\!=\!1$ and $V\!=\!1$, which are the normalized diameter and boundary velocity, respectively. The symbol $\theta (^\circ)$ denotes the angle, as shown in Fig. \ref{fig-5}. We find that the simulation data are in good agreement with the theoretical prediction. Interestingly, $V_z(r,\theta)$ is not influenced by the boundary velocity rotation. Moreover, as shown below, the Brownian force ($\Leftrightarrow$ nonzero $D$) has no influence on $V_z(r,\theta)$.

The results obtained for lattice B are shown in Figs. \ref{fig-6}(d)--(f). Distributions $h(V_z)$ and $h(V)$ in Figs. \ref{fig-6}(d) and (e) are almost the same because the boundary velocity is along the $z$ direction on lattice B. By contrast, they are slightly different from each other, as shown in Fig. \ref{fig-6}(a) and (b) on lattice A, as expected from the boundary velocity rotating around the cylinder. It is noteworthy that $V_z^{\rm max}(=\!41.5)$ on lattice A is smaller than $V_z^{\rm max}(=\!46.6)$ on lattice B, and moreover, that $h(V_z)$ on lattice A is slightly smaller than that on lattice B for all $V_z\!<\!40 ({\rm \mu m/s})$ in Figs. \ref{fig-6}(a), (d) as indicated by the updown arrows at $V_z\!=\!25 ({\rm \mu m/s})$. By contrast, $h(V)$ on lattice A is apparently larger than that on lattice B for all $V$ except for the velocities at $h(V)\!=\!1$ in Figs. \ref{fig-6}(b), (e). This difference is nontrivial and implies that the rotating fluid circulation enhances the flow velocity inside, corresponding to the mixing enhancement \cite{Goldstein-etal-PRL2008,Goldstein-etal-PNAS2008,Goldstein-etal-JFM2010,Raymond-Goldstein-IF2015}. The $r$ dependence of $V_z(r,\theta)$ in Fig. \ref{fig-6}(c) are the same as those in Fig. \ref{fig-6}(f) and almost the same as those reported in Ref. \cite{Goldstein-etal-JFM2010}. Note that the boundary velocities are not included in the calculations of $h(V_z)$, $h(V)$ and $V_z(r,\theta)$.

\subsection{Velocity distribution for $D\!\not=\!0$ \label{D_nonzero}}
\begin{figure}[t]
\begin{center}
\includegraphics[width=11.0cm]{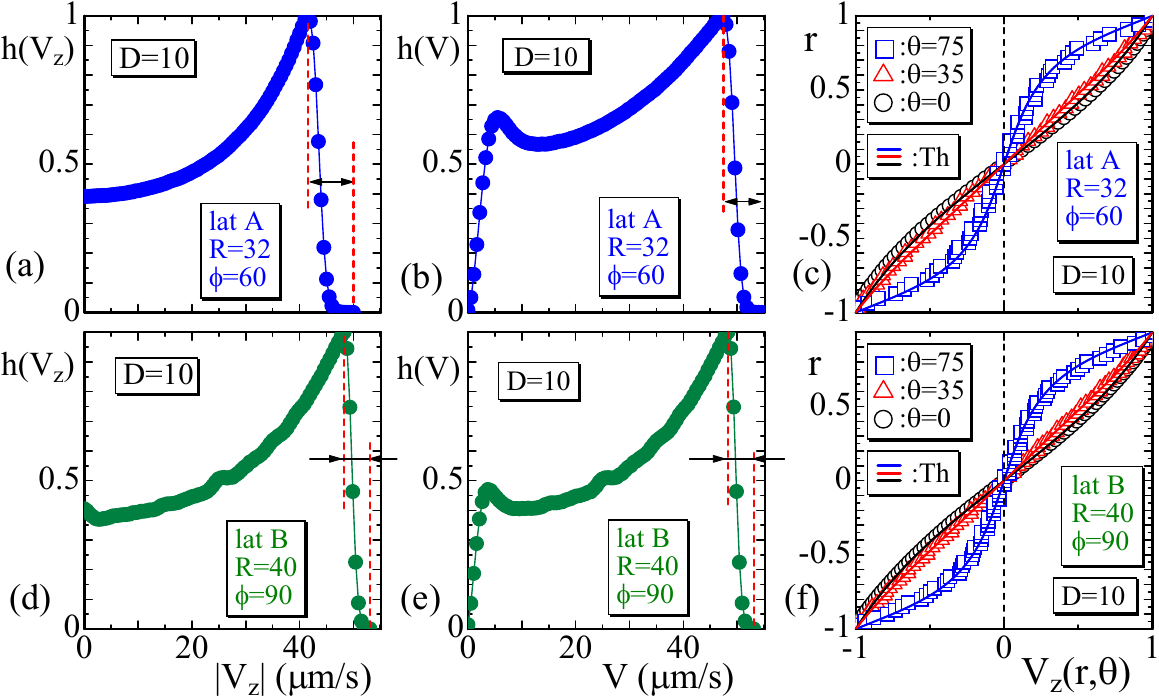}
\caption{ (a) $h(V_z)$ vs. $|V_z|$, (b) $h(V)$ vs. $V$, and (c) radial dependence of normalized $V_z(r,\theta)$ obtained on lattice A, and (d), (e), and (f) those obtained on lattice B. $D$ is fixed to $D\!=\!10$ in the simulation units. The solid lines denoted as Th in (c) and (f) are obtained from Eq. (\ref{Pickard}). 
 \label{fig-7} }
\end{center}
\end{figure}
\begin{figure}[t]
	\begin{center}
		\includegraphics[width=11.05cm]{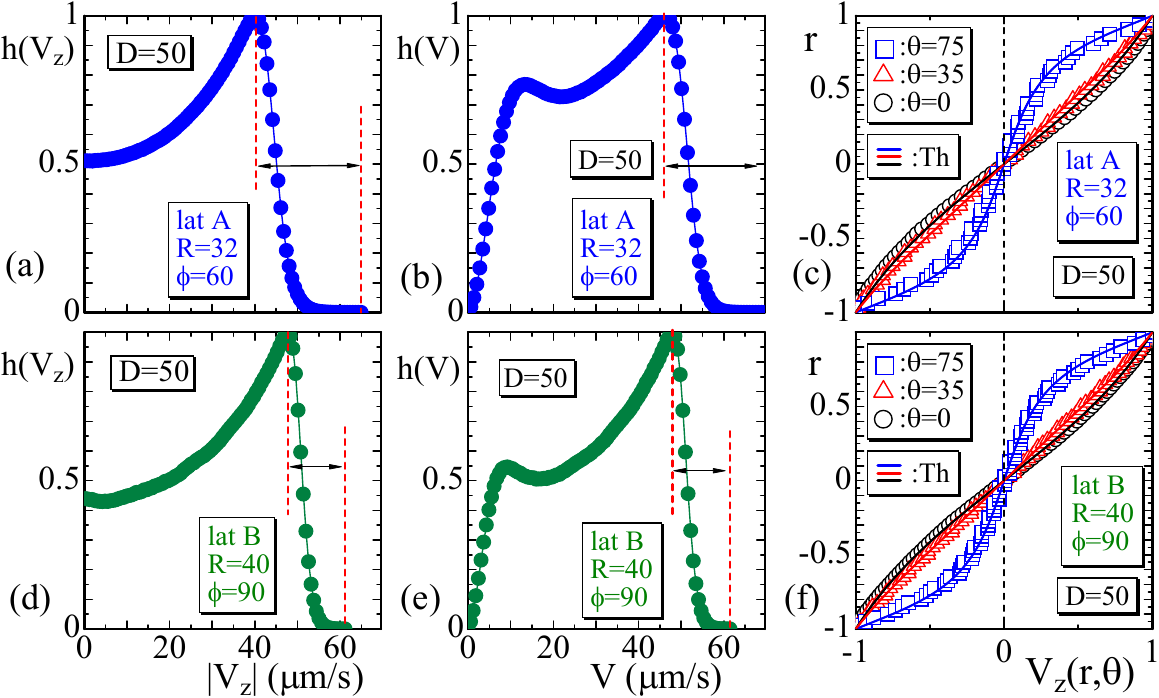}
		\caption{(a) $h(V_z)$ vs. $|V_z|$, (b) $h(V)$ vs. $V$, and (c) radial dependence of normalized obtained on lattice A; (d), (e), and (f) those obtained on lattice B. $D$ is fixed to $D\!=\!50$ in the simulation units. 
			The solid lines denoted as Th in (c) and (f) are obtained from Eq. (\ref{Pickard}).
			\label{fig-8} }
	\end{center}
\end{figure}
\begin{figure}[t]
	\begin{center}
		\includegraphics[width=11.0cm]{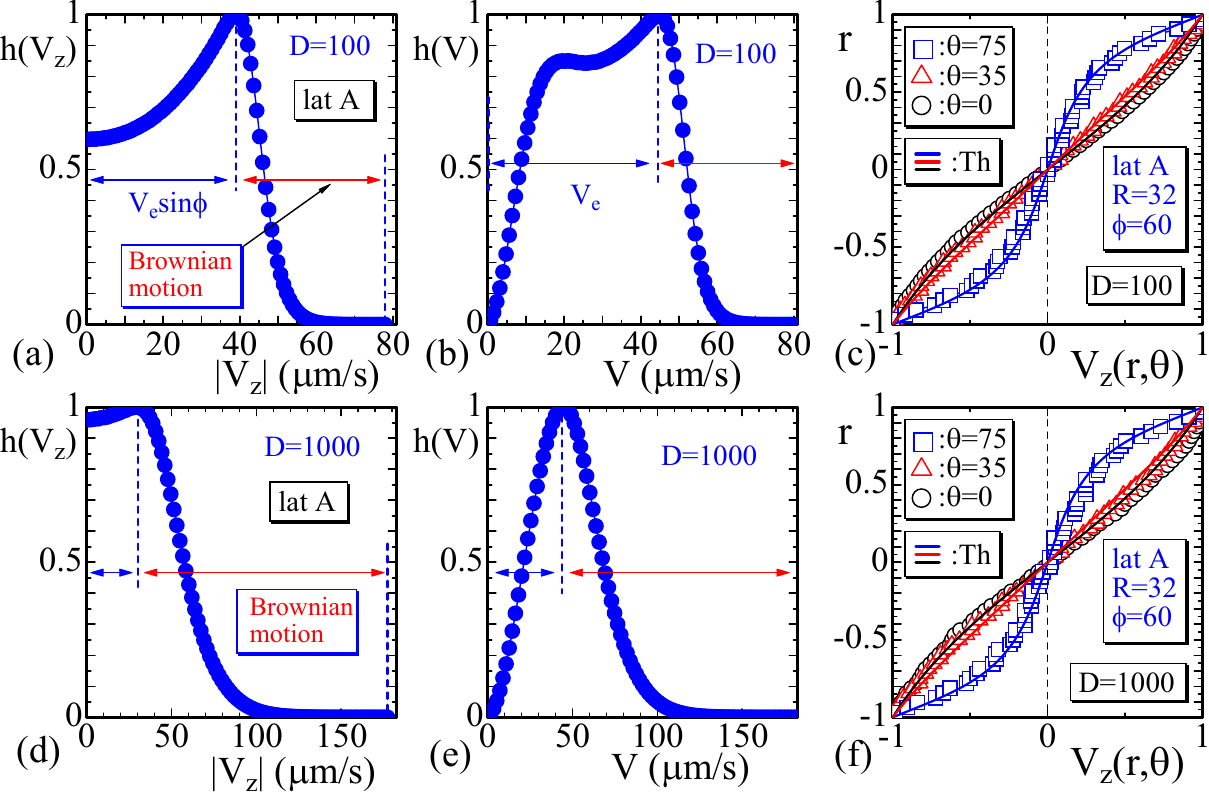}
		\caption {$h(V_z)$ vs. $|V_z|$, $h(V)$ vs. $V$, and radial dependence of normalized $V_z(r,\theta)$ at (a)--(c) $D\!=\!100$, and (d)--(f) $D\!=\!1000$ obtained on lattice A. The peak velocities $V_z^{\rm peak}$ and $V^{\rm peak}$ are slightly smaller than the the boundary velocities $V_e\sin\phi\!=\!50 \sin 60^\circ\!\simeq\!43 ({\rm \mu m})$ and $V_e\!=\!50 ({\rm \mu m})$. Enhancements of velocities, denoted by ``Brownian motion'', become apparent with increasing $D$.
			\label{fig-9} }
	\end{center}
\end{figure}
The results corresponding to the $D\!=\!10$ are plotted in Figs. \ref{fig-7}(a)--(c) and \ref{fig-7}(d)--(f) for lattices A and B, respectively. We find that $h(V_z)$ and $h(V)$ are different from those plotted in Fig. \ref{fig-6} for both lattices A and B. Indeed, $h(V_z)$ and $h(V)$ drop to zero at $V\!\simeq\! 45 ({\rm \mu m/s})$ in Fig. \ref{fig-7}. This implies that there are fluid velocities $V_z\!>\!V_e \sin \phi^\circ$ and $V\!>\!V_e\!=\!50 ({\rm \mu m/s})$, where $\phi=\!60^\circ (90^\circ)$ on lattice A (lattice B). The reason for the appearance of $V_z\!>\!V_e \sin \phi^\circ$ and $V\!>\!V_e$, indicated by left-right arrows ($\leftrightarrow$, $\rightarrow\leftarrow$), is that the velocity is activated by Brownian forces. Owing to this nontrivial contribution, a tail of the velocity appears in $h(V_z)$ and $h(V)$ in the regions of $V_z\!>\!V_e \sin \phi$ and $V\!>\!V_e$. The drop in $h(V)$ at $V\!\simeq\! 10 ({\rm \mu m/s})$ is also a nontrivial influence of Brownian force. The  $r$-dependence of $V_z(r,\theta)$ is independent of lattices A and B, as shown in Fig. \ref{fig-7}(c) and (f).

Next, we plot the results corresponding to $D\!=\!50$ for lattices A and B in Figs. \ref{fig-8}(a)--(c) and  \ref{fig-8}(d)--(f), respectively. The peak positions of $V^{\rm peak}$ for $h(V)$ in Fig. \ref{fig-8}(b) on lattice A are almost the same as those in Fig. \ref{fig-8}(e) on lattice B. In contrast, the velocity regions larger than the peaks; $V_z\!>\!V_z^{\rm peak}$ and $V\!>\!V^{\rm peak}$ in Figs. \ref{fig-8}(a),(b) indicated by $(\leftrightarrow)$, are wide compared to those in Figs. \ref{fig-8}(d),(e) implying that the rotating fluid circulation is effective for velocity enhancement if the Brownian force is increased.  The radial dependence of $V_z(r,\theta)$ is almost the same for lattices A and B, even at $D\!=\!50$, as shown in Fig. \ref{fig-8}(c) and (f).

To observe the dependence on $D$, we further increase $D$ to $D\!=\!100$ and $D\!=\!1000$ for $h(V_z)$ and $h(V)$, as shown in Figs. \ref{fig-9}(a)--(f), where only the results for lattice A are presented. In the larger $D$ regions, the effects of Brownian forces on the enhancement of flow velocity are apparent. In the figures, this enhancement is denoted by left-right arrows ($\leftrightarrow$) with the symbol ``Brownian motion.” We should emphasize that the radial dependence of $V_z(r,\theta)$ remains unchanged, even for sufficiently large $D$ such as $D\!=\!1000$ as plotted in Figs. \ref{fig-9}(c),(f). However, this is reasonable because the fluid particles at position $(r,\theta)$ thermally fluctuate in isotropic directions with rapid velocity, and the mean value $V_z(r,\theta)$ has no specific direction dependence; therefore, the shape of the curves $V_z(r,\theta)$ remains unchanged.

\subsection{Velocity distribution considering $D$ variation \label{peak_velocity}}
\begin{figure}[ht]
\begin{center}
\includegraphics[width=11.0cm]{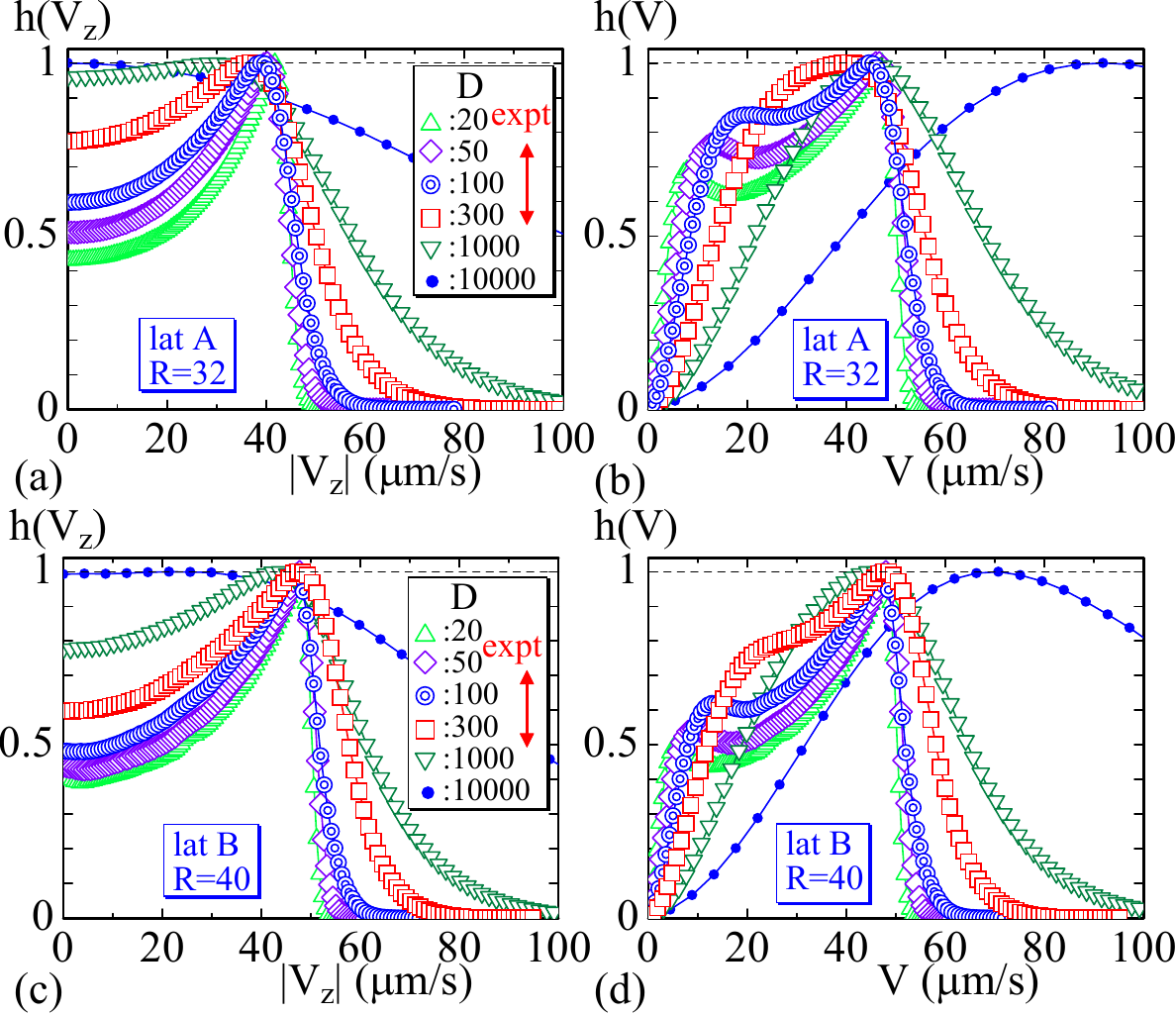}
\caption{(a) $h(V_z)$ vs. $|V_z|$ and (b) $h(V)$ vs. $V$ obtained for $20\!\leq\!D\!\leq\!10^4$ on lattice A. (c), (d) those obtained  on lattice B. The peaks are approximately $V_z^{\rm peak}\!=\!40 ({\rm \mu m/s})$ in (a) and $V^{\rm peak}\!=\!46 ({\rm \mu m/s})$ in (b). These values are slightly smaller than the boundary velocities $43(=\!V_e\sin60^\circ) ({\rm \mu m/s})$ and 50 (${\rm \mu m/s}$), respectively. The up-down arrows with ``expt'' in (a) and (c) denote the experimentally relevant region, in the sense that the shape of $h(V_z)$ is close to the experimentally observed shapes in Figs. \ref{fig-2}(c), (d). \label{fig-10} }
\end{center}
\end{figure}

In this subsection, we analyze the dependence of the peak velocities on $D$ and discuss whether the peak positions are influenced by the thermal fluctuations of $\vec{V}$ to find $D$ which is suitable for the system under consideration in the sense that the shape of $h(V_z)$ is close to the experimental ones in Figs. \ref{fig-2}(c), (d). In addition, by finding $D$ or a range of $D$ such that the shape of $h(V_z)$ is close to the experimental shape, we show that the LNS simulation in this study is suitable for studying protoplasmic streaming, as mentioned in the final part of the preceding section.

Figures \ref{fig-10}(a) and (b) show $h(V_z)$ versus $|V_z|$ and $h(V)$ versus $V$ obtained in the range of $20\!\leq\!D\!\leq\!10^4$ on lattice A. The peak positions in Figs. \ref{fig-10}(a),(b) are almost independent of $D$, and these are close to or slightly smaller than $V_e\cos 60^\circ\!\simeq\!43 ({\rm \mu m/s})$ and $V_e\!=\!50({\rm \mu m/s})$, respectively, in the range of $D\!\leq\!1000$ at least. We find an experimentally relevant region of $D$, as indicated by the updown arrow in Figs. \ref{fig-10}(a),(b), where the shape of $h(V_z)$ is close to the experimental ones in Figs. \ref{fig-2}(c), (d).

We plot the $h(V_z)$ and $h(V)$ results obtained on lattice B in Figs. \ref{fig-10}(c),(d). The  peak positions of $h(V_z)$ and $h(V)$ in Figs. \ref{fig-10}(c), (d), are the same because the boundary velocity is parallel to the $z$ axis on lattice B. These peaks are approximately equal to $V_z^{\rm peak}\!=\!V^{\rm peak}\!=\!47 ({\rm \mu m/s})$ for $D\!\leq\!300$ in Figs. \ref{fig-10}(c), (d). We find that the shape of $h(V_z)$ at $D\!=\!300$ is relatively close to the experimentally observed shape, and hence, the experimentally relevant region of $D$, indicated by the updown arrow in Fig. \ref{fig-10}(c), moves to slightly larger region compared to that on lattice A in Fig. \ref{fig-10}(a).

$V^{\rm peak}$ and $V_z^{\rm peak}$ are listed in Table \ref{table-4} with $D$ and the corresponding $D_e$ in the physical unit, which represent the strength of the numerically introduced  single Brownian force defined by Eq. (\ref{Brownian-motion-dt}). $D_{e,\tau}$ corresponds to the strength of the random force in the case ${\it \Delta}\tau\!=\!5\!\times\!10^{-4} ({\rm s})$ and is obtained by the formula $D_{e,\tau}\!=\!D_{e}\frac{{\it \Delta}\tau}{{\it \Delta}t}$ in Eq. (\ref{D-correspondence}).  Strengths $D_{e,\tau}$ in the range of $50\!\leq\!D\!\leq 100$ are consistent with  Eq. (\ref{D-vs-nu}), as expected from the fluctuation-dissipation relation, as discussed in Appendix \ref{append-D}.
			\begin{table}[h!]
\caption{
\noindent
	Assumed parameters $D$,  $D_{e}$, $D_{e,\tau}$ in the simulation unit and physical unit at  ${\it \Delta}\tau\!=\!{\it \Delta}t\!=\!2\!\times\!10^{-8}({\rm s})$ and  ${\it \Delta}\tau\!=\!5\!\times\!10^{-4} ({\rm s})$, the peak values $V_z^{\rm peak}$ and $V^{\rm peak}$ of $h(V_z)$ and $h(V)$ on lattice A in Figs. \ref{fig-10}(a), (b).
\label{table-4}}
\begin{center}
 \begin{tabular}{cccccccccc }
 \hline 
$D$ && $D_{e}$ &&  $D_{e,\tau}$ && $V_z^{\rm peak} ({\rm lat\, A})$&& $V^{\rm peak} ({\rm lat\, A})$ \\
$\frac{\rm (\alpha_0 m)^2}{({\rm \beta_0 s})^3}$ && ${\rm m^2}/{\rm s}^3$&& ${\rm m^2}/{\rm s}^3$ && ${\rm \mu m/s}$&& ${\rm \mu m/s}$ \\
 \hline 
 20 && $1.25\times10^{-6}$ && $3.1\times10^{-2}$ &&42.2 && 47.6 \\
 50 && $3.13\times10^{-6}$ && $7.8\times10^{-2}$ &&40.7 && 46.4 \\
 100 && $6.25\times10^{-6}$  && $1.6\times10^{-1}$ &&39.4 && 45.0 \\
 300 && $1.88\times10^{-5}$  && $4.7\times10^{-1}$ &&37.2 && 40.8 \\
 1000 && $6.25\times10^{-5}$ && $1.6$ &&30.4 && 44.3 \\
 3000 && $1.88\times10^{-4}$ && $4.7$ &&9.05 && 58.4 \\
\hline
\end{tabular}
\end{center}
\end{table}

\subsection{Theoretical prediction of the velocity distribution \label{theoretical_prediction}}
\begin{figure}[ht]
	\begin{center}
		\includegraphics[width=13.5cm]{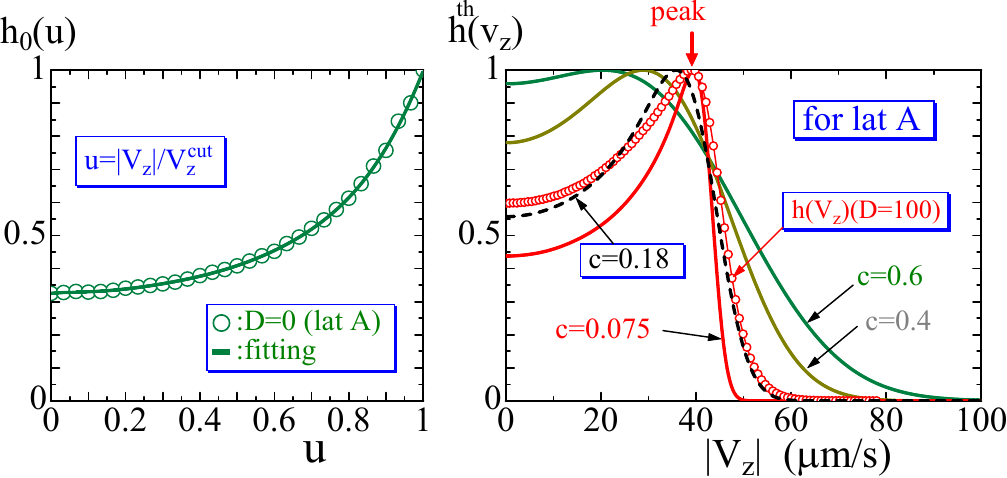}
		\caption{(a) Plotted circles (\textcolor{green}{$\bigcirc$}) are obtained by normalizing the simulation results of $h(V_z)$ at $D\!=\!0$ in Fig. \ref{fig-6}(a) by $u=|V_z|/V_z^{\rm cut}$ with $V_z^{\rm cut}\!=\!41.5 ({\rm \mu m/s})$ and the corresponding solid line is drawn by fitting with a polynomial function $h_0(u)$ (Appendix \ref{append-E}). 
			(b) $h^{\rm th}(v_z)$ vs. $|V_z|(=41.5|v_z|) ({\rm \mu m/s})$. We find that the curve with $c\!=\!0.18$ is relatively close to the simulation data ($\textcolor{red}{\circ}$) at $D\!=\!100$, which is an experimentally relevant value. 
		\label{fig-11} }
	\end{center}
\end{figure}
In this study, we simulate the Brownian motion of biological materials by using the LNS equation in Eq.(\ref{NS-eq-org}) for fluids without biological materials. For this reason, it is meaningful to compare the simulation results with the Maxwell-Boltzmann distribution of the velocity $V_z$ of the Brownian particles. To verify that the simulation results are reasonable, we assume that the Brownian particles flow with the same velocity distribution $h(V_z)$ as that of the fluid in Fig. \ref{fig-6}(a) for $D\!=\!0$. We calculate a theoretical distribution $h^{\rm th}(v_z)$ of the particles by 
\begin{eqnarray}
	\label{theoretical}
	\begin{split}
		&h^{\rm th}(v_z)\propto \int_{-1}^{1} h_0(u) \exp \left[-(v_z-u)^2/c^2\right]du, \\
		& h_0(u)={\rm polynomial\; function},\\
		& \left(u=1 \Leftrightarrow V_z=41.5 ({\rm \mu m/s})\right), 
	\end{split}
\end{eqnarray}
where $h_0$ is the polynomial function of $u(=\!V_z/41.5)$, which is a normalized velocity (Appendix \ref{append-E}), fitting the simulation result $h(u)$ on lattice A at $D\!=\!0$ corresponding to $h(V_z)$ in Fig. \ref{fig-6}(a), where the maximum velocity $V_z^{\rm cut}\!=\!41.5 ({\rm \mu m/s})$ is indicated by the arrow. This $h_0(u)$ represents the effect of the boundary flow. The meaning of $\exp \left[-(v_z-u)^2/c^2\right]$, denoted by $f(v_z,u)$, is the Maxwell-Boltzmann distribution of the fluctuating velocity $v_z$ of particles moving with velocity $u$. In the expression of $h^{\rm th}(v_z)$, the unit velocity is assumed to be $v_z=1 \Leftrightarrow V_z=41.5 ({\rm \mu m/s})$. The constant $c$ in the exponential function is the peak position of the Maxwell-Boltzmann distribution $f(v) \propto v^2\exp(-v^2/c^2)$. The reasons for the multiplication of $h_0(u)$ and the replacement $v_z^2$ in $f(v_z,u\!=\!0)$ to $(v_z\!-\!u)^2$ in $f(v_z,u)$ are to include effects of Brownian motion of all moving particles with mean velocity $|u|\!\leq\!1$ into the distribution $h^{\rm th}(v_z)$, in which $v_z$ is not always limited to $|v_z|\!\leq\!1$. Specifically, the expression $h_0(u)du$ for $|u|\!\leq\!1$ is proportional to the total number of particles with mean velocity between $u$ and $u\!+\!du$, and $\exp \left[-(v_z-u)^2/c^2\right]$ is the contribution of the moving particles of velocity $u$ to $ h^{\rm th}(v_z)$, and these contributions can be integrated out in the range $-1\!\leq\!u\!\leq\!1$. In this calculation, it is assumed that the particles move with velocities of distribution $h_0(u)$ along the $z$ axis; hence, no interaction between the particles and fluids is assumed.

Figure \ref{fig-11}(a) shows $h_0(u)$  versus  $u(=\!|V_z|/V_z^{\rm cut})$, where $V_z^{\rm cut}\!=\!41.5({\rm \mu m/s})$. $h^{\rm th}(v_z)$ is plotted with physical unit of $|V_z|$ in Fig. \ref{fig-11}(b), where $V_z\!=\!V_z^{\rm cut}v_z$. Simulation data obtained at $D\!=\!100$ plotted with the symbol ($\textcolor{red}{\circ}$) are close to the dashed line at $c\!=\!0.18$. This indicates that the velocity distribution of Brownian particles is suitably reflected in the simulation data.

\subsection{Mean velocities and mixing enhancement	\label{momentum_conservation}}
\begin{figure}[ht]
\begin{center}
\includegraphics[width=10.5cm]{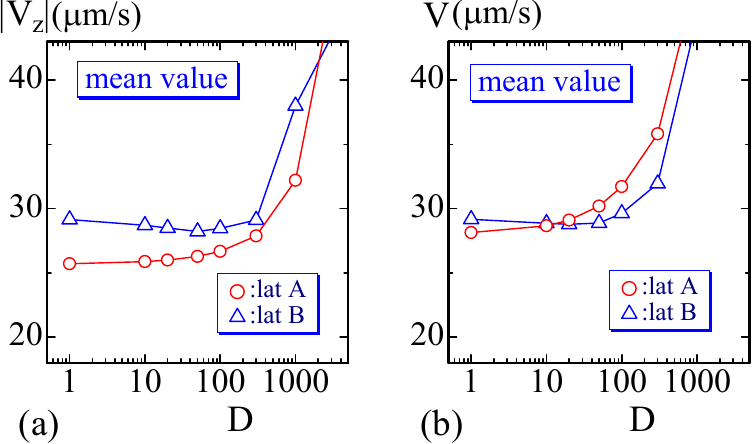}
\caption{The mean values of (a) $|V_z|$ vs. $D$ and (b) $V$ vs. $D$ obtained on both lattices A and B. The mean value $V$ of lat A becomes larger than that of lat B at the experimentally relevant  $D (\sim 100)$ region.
\label{fig-12} }
\end{center}
\end{figure}

 The mean values of $|V_z|$ and $V$ are shown in Figs. \ref{fig-12}(a),(b). The mean values are calculated using Eq. (\ref{mean-of-Q}), in which each sample is obtained from the lattice averages of $|V_z|$ and $V$ such that $|V_z|\!=\!\sum_{i}|V^z_{i}|/\sum_{i}1$ and $V\!=\!\sum_{i}\|\vec{V}_{i}\|/\sum_{i}1$.  The convergence of the time-evolution iterations of the lattice averages $|V_z|$ and $V$ is presented in Appendix \ref{append-F}. The total number of samples is  $n_{\rm s}\!=\!100$, which is smaller than  $n_{\rm s}\!=\!1000$ for the calculation of $\vec{V}(r,\theta)$ at a lattice section (Fig. \ref{fig-5}), however, the data $|V_z|$ and $V$ plotted in Figs.  \ref{fig-12}(a),(b) are calculated on the whole lattice points except the boundary $\Gamma_3$ (Fig. \ref{fig-3}), and therefore, the statistics is considered to be sufficient. We find from Fig. \ref{fig-12}(b) that 
\begin{eqnarray}
	\label{mixing-enhancement}
	V ({\rm lat \; A}) > V  ({\rm lat \; B}), \quad (D\geq 50)
\end{eqnarray}
is satisfied in the region of $D\geq50$ including the experimentally relevant $D$ region, 
whereas, $|V_z| ({\rm lat \; A}) < |V_z|  ({\rm lat \; B}) \; (D < 3000)$. The result of Eq. (\ref{mixing-enhancement}) indicates that the mixing along the $\vec{V}$ direction is enhanced by the Brownian motion under the boundary fluid circulation.

\subsection{Snapshots of the velocity and pressure \label{velocity-pressure}}
\begin{figure}[h!]
\begin{center}
\includegraphics[width=12.0cm]{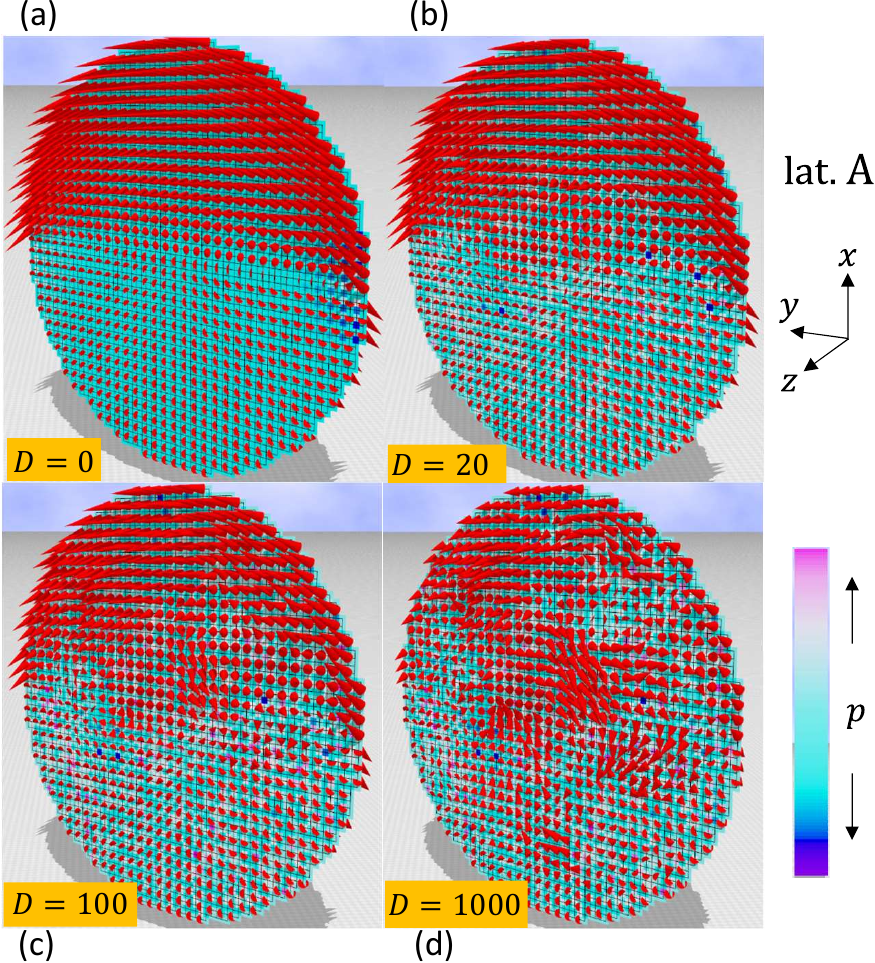}
\caption{Snapshots of the velocity and pressure obtained on lattice A at the cross-section at the middle of the cylinder $z\!=\!L/2(\simeq\!179)$ corresponding to  four different $D$. The small red cones represent the velocity. Only the velocities at every other vertex are shown.
 \label{fig-13} }
\end{center}
\end{figure}
We show snapshots of the velocity and pressure obtained on the cross-section of the cylinder at $z\!=\!L/2$ on lattice A for several different $D$ values. The direction of the boundary velocity at $z\!=\!L/2$ is the opposite to that on the boundaries $\Gamma_1$ at $z\!=\!0$ and $\Gamma_2$ at $z\!=\!L$ (Fig. \ref{fig-3}(a)). The velocities denoted by the cones in Figs. \ref{fig-13}(a)--(d) are of convergent configurations corresponding to (a) $D\!=\!0$, (b) $D\!=\!20$, (c) $D\!=\!100$ and (d) $D\!=\!1000$. For a clear visualization, only the velocities at every other vertex are shown. The pressures $p$ are normalized to $0\!\leq\!p\!\leq\!1$ and represented by the color gradation. In this normalization, the boundary pressure $p\!=\!0$ changes to $p\!\simeq\!0.5$. Because $V_z\!<\!0$ in the lower part of the cross-section, the velocities are hidden behind the cross-sectional surfaces for the pressure visualization.

\begin{figure}[h!]
\begin{center}
\includegraphics[width=12.0cm]{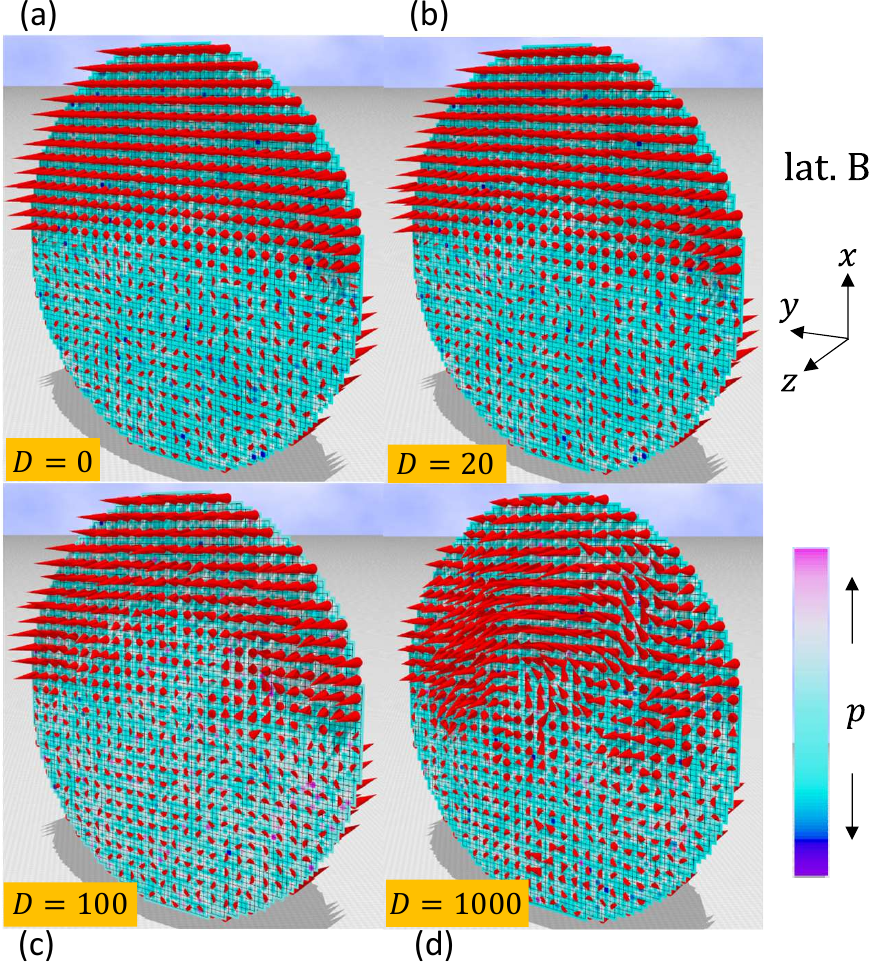}
\caption{Snapshots of velocity and pressure obtained on lattice B at the cross-section of the middle of the cylinder $z\!=\!L/2(=\!40)$ corresponding to four different $D$. The small red cones represent the velocity. Only the velocities at every third vertex are shown.  \label{fig-14} }
\end{center}
\end{figure}
An examination of Fig. \ref{fig-13}(a) shows that the fluid regularly flows according to the boundary velocity, and the pressure $p$ remains almost unchanged from the boundary pressure $p(\simeq\!0.5)$ at $D\!=\!0$. The velocity and pressure are confirmed to be disturbed at nonzero $D$, and the disturbance becomes stronger when $D$ increases.

The pressure at higher $D$ does not always vary smoothly but is randomly distributed over the cross-section. This condition of the pressure configuration is relatively close to that of Couette flow between parallel plates, which was obtained using 2D LNS simulations \cite{Noro-etal-2021}.

To visualize the difference in velocity configuration $\vec{V}$ between lattices A and B, we show snapshots of  $\vec{V}$ and pressure $p$ obtained on lattice B in Figs. \ref{fig-14}(a)--(d). The velocities in Figs. \ref{fig-14}(a)--(d) are shown at every third vertex. Because $p\simeq\!0.5$ at every point on the cross-section at $D\!=\!0$ in the convergent configuration, we show snapshots of $\vec{V}$ and $p$ at $D\!=\!1$ instead of those at $D\!=\!0$ in Fig. \ref{fig-14}(a). The velocity configurations on lattice B are clearly different from those on lattice A for all $D$, as expected from the difference in the boundary condition for $\vec{V}$.

\section{Summary and conclusion\label{summary}}
In this paper, we numerically study the velocity distribution for protoplasmic streaming in plant cells using Langevin and Navier-Stokes (LNS) simulation on two different 3D lattices, A and B. The boundary velocity rotates on the cylindrical surface of lattice A, whereas it is parallel to the longitudinal direction on lattice B. In the LNS simulation scheme, the velocity field $\vec{V}$ is treated as a fluctuating stochastic variable under a Brownian random force of strength $D$. Therefore, the simulated velocity distributions can be compared and identified with the reported experimentally observed velocity distribution. The goal of this study is to confirm this expectation.  Additionally, we are interested in whether the velocity fluctuations due to Brownian forces have nontrivial effects on the mixing of flows inside the cells.

We find that the velocity distribution $h(V_z)$ of $|V_z|$ has a peak corresponding to the boundary velocity for $D\!\geq\!10$. The shape of $h(V_z)$ on lattice A in the range $50\!\leq\!D\!\leq 1000$ is close to the reported experimental data. Moreover, the $D$ value is predicted to be in the region $50\!\leq\!D\!\leq 100$ from the fluctuation dissipation relation of Landau and Lifschitz. The radial dependence of $\vec{V_z}$ calculated at several angles $\theta$ from the vertical direction  on the cross section is in good agreement with the theoretical prediction.

We also find two different peaks in $h(V)$, $(V\!=\!\|\vec{V}\|)$ at two different velocities $V_1$ and $V_2$ ($V_1<V_2$) when $D$ is increased to $10\!\leq\!D\!\leq 100$ or greater, where the second peak $V_2$ corresponds to the boundary velocity. The Brownian force is reflected in the emergence of a tail in $h(V)$ for $V>V_2$. This appearance of  velocities $V>V_2$ higher than the boundary velocity $V_2$ is expected to play a nontrivial role in enhancing mixing. 

The simulated velocity distribution $h(V_z)$ is also compared with the theoretical distribution of Brownian particles moving with  velocity $u$ and distribution $h(u)$, which is the same as the simulated  $h(V_z)$ of the streaming with $D\!=\!0$. By assuming a Maxwell-Boltzmann distribution for the velocity of moving particles, we confirm that the obtained theoretical prediction is in good agreement with the simulation data.  This consistency implies that the LNS equation used in this study suitably describes the Brownian motion of biological materials.

To observe the enhancement of mixing, we calculate the mean values of $V_z$ and  $\vec{V}$ on both lattices A and B by varying $D$. From these calculations, we find that the mixing of biological materials is enhanced along the direction of  $\vec{V}$ by the Brownian motion of biological materials in the experimentally relevant region of $D$ on lattice A with rotating circular boundary fluids. Importantly, mixing is improved in the presence of the Brownian motion of biological materials. In addition, using the mean values of $V_z$ and  $\vec{V}$, we confirm the momentum conservation, which can be a numerical check for the correspondence between the Gaussian random force in this study and the fluid dynamical random force of Landau-Lifshitz.

\acknowledgements
H.K. acknowledges Dr. Fumitake Kato for discussions. This work was supported in part by Collaborative Research Project J20Ly09 at the Institute of Fluid Science (IFS), Tohoku University, and in part by a Collaborative Research Project at the National Institute of Technology (KOSEN), Sendai College. Numerical simulations were performed on the "AFI-NITY" supercomputer system at the Advanced Fluid Information Research Center, Institute of Fluid Science, Tohoku University.

\appendix

\section{SMAC method, convergence criteria and the divergenceless condition \label{append-A}}
In this Appendix, we show how to obtain the solution $\vec{V}(t)$ in Eq. (\ref{NS-eq-time-step}) under the condition ${\partial {\vec V}}/{\partial t}\!=\!0$ in Eq. (\ref{SS-condition}). The relationship between Gaussian random number $\vec{g}$ and Brownian force $\vec{\eta}$ is given by $\sqrt{2D{\it \Delta} t} \,\vec{g}\!=\!\vec{\eta}{\it \Delta} t$ \cite{Egorov-etal-POF2020}.
Note that the discrete time $t$ in Eq. (\ref{NS-eq-time-step}) is introduced to obtain the steady-state solution that satisfies Eq. (\ref{SS-condition}) and is different from the real time $t$ in Eq. (\ref{NS-eq-org}).
This time can be called ``fictitious time'' because the divergenceless condition of Eq. (\ref{SS-condition}) is not always satisfied until a convergent solution corresponding to the random force $\sqrt{2D/{\it \Delta} t}\,\vec{g}$ is obtained.
Thus, the Brownian random force $\vec{g}$ is incremented only when the convergent solution $\vec{V}$ of Eq. (\ref{NS-eq-time-step}) is obtained. 
From Eq. (\ref{NS-eq-time-step}), we understand that $\nabla\cdot {\vec V}(t+{\it \Delta} t)\!=\!0$ is not always satisfied even if $\nabla\cdot {\vec V}(t)\!=\!0$ is satisfied because the terms independent of ${\vec V}(t)$ on the right-hand side are not always divergenceless. Moreover, the time evolution of $p(t)$ is not specified. Therefore, we introduce a temporal velocity $\vec{V}^*(t)$ and rewrite Eq. (\ref{NS-eq-time-step}) as follows:
\begin{eqnarray}
\label{NS-eq-time-step-temporal-1}
&&{\vec V}^*(t)= \vec{V}(t) +{\it \Delta} t \left[ \left (-{\vec V}\cdot \nabla\right){\vec V}(t)-{\rho}^{-1} {\it \nabla} p(t) +\nu \Laplace {\vec V}(t)\right] +\sqrt{2D{\it \Delta} t} \,\vec{g}(t),\\
\label{NS-eq-time-step-temporal-2}
&&{\vec V}(t+{\it \Delta} t)={\vec V}^*(t)-{\it \Delta} t {\rho}^{-1} {\it \nabla} \left[ p(t+{\it \Delta} t)-p(t)\right].
\end{eqnarray}

By applying the divergence operator $\nabla\cdot$ to Eq. (\ref{NS-eq-time-step-temporal-2}),
we obtain
\begin{eqnarray}
\label{SMac-eq-0}
\nabla\cdot{\vec V}(t+{\it \Delta} t)=\nabla\cdot{\vec V}^*(t)-{\it \Delta} t {\rho}^{-1} \Laplace \left[ p(t+{\it \Delta} t)-p(t)\right].
\end{eqnarray}
Then, assuming the condition $\nabla\cdot{\vec V}(t+{\it \Delta} t)\!=\!0$,
we obtain Poisson's equation for $\phi(t) \!=\! p(t+{\it \Delta} t)\!-\!p(t)$:
\begin{eqnarray}
\label{SMac-Poiss}
\Laplace \phi(t)=\frac{\rho}{{\it \Delta} t}\nabla\cdot{\vec V}^*(t),\quad \phi(t) = p(t+{\it \Delta} t)-p(t).
\end{eqnarray}
Thus, combining Eq. (\ref{NS-eq-time-step-temporal-1}) for the time evolution of ${\vec V}^*(t)$ with Poisson's equation in Eq. (\ref{SMac-Poiss}) for $\phi(t) \!=\! p(t+{\it \Delta} t)\!-\!p(t)$, we implicitly obtain the time evolution ${\vec V}(t+{\it \Delta} t)$ with the condition $\nabla\cdot{\vec V}(t+{\it \Delta} t)\!=\!0$. The time evolution of $p$ from $p(t)$ to $p(t\!+\!{\it \Delta} t)$ can also be obtained by adding the solution $\phi(t)$ to $p(t)$, i.e., $p(t)\!+\!\phi(t)$. In this technique, an explicit time-evolution step is assumed for $\vec{V}(t)$, whereas an implicit time-evolution is assumed for $p(t)$ by solving the Poisson equation for $\phi(t)$. The discrete time step ${\it \Delta} t$, which is given in Table \ref{table-3}, is fixed, independent of $D$.  For fast convergence, a large ${\it \Delta} t$ is preferable. However,  ${\it \Delta} t$ should be sufficiently small for a combination with a sufficiently large $D$ such as $D\!=\!10000$. Note that if ${\it \Delta} t$ is changed, $D$ should also be changed to obtain the same results \cite{Egorov-etal-POF2020,Noro-etal-2021}.

The simulation procedure can be summarized as follows:
\begin{enumerate}
\item[(i)] Calculate $V^*(t)$ by Eq. (\ref{NS-eq-time-step-temporal-1}) using the current $V(t)$, $p(t)$ and $\vec{g}$
\item[(ii)] Solve Poisson's equation for $\phi(t)$ in Eq. (\ref{SMac-Poiss})
\item[(iii)] Calculate $V(t+{\it \Delta} t)$ and $p(t+{\it \Delta} t)$ by Eq. (\ref{NS-eq-time-step-temporal-2}) and $p(t)\!+\!\phi(t)$, respectively
\item[(iv)] Repeat steps (i)--(iii) until the convergence criteria given below are satisfied
\end{enumerate}
This technique for updating ${\vec V} (t)$ is slightly different from the original MAC method, where ${\vec V} (t)$ is explicitly updated to ${\vec V}(t+{\it \Delta} t)$. Hence, $\nabla\cdot{\vec V}\!=\!0$ is not always satisfied and may be slightly violated even for the convergent solution. This violation becomes larger for a larger Brownian force strength $D$ in the original MAC method; however, it is negligibly small for the convergent solutions in the SMAC method. Detailed information on $\nabla\cdot{\vec V}\!=\!0$ is given below. The most time-consuming part of this process is to solve the Poisson equation for $\phi$, which is simulated using the Open-Mp parallelization technique coded in Fortran.

We assume the following convergence criteria for $\vec{V}$ and $p$:
\begin{eqnarray}
\label{convergent-time-step}
\begin{split}
&{\rm Max}\left[\left||\nabla\cdot\vec{V}_{ijk}(t+{\it \Delta} t)|-|\nabla\cdot\vec{V}_{ijk}(t)|\right|\right]<1\times 10^{-8},\\
&{\rm Max}\left[|\vec{V}_{ijk}(t+{\it \Delta} t)-\vec{V}_{ijk}(t)|\right]<1\times 10^{-8},\\
&{\rm Max}\left[|p_{ijk}(t+{\it \Delta} t)-p_{ijk}(t)|\right]<1\times 10^{-8},
\end{split}
\end{eqnarray}
and the criterion for the Poisson equation iterations is
\begin{eqnarray}
\label{convergent-Poisson}
{\rm Max}\left[|\phi_{ijk}(n+1)-\phi_{ijk}(n)|\right]<1\times 10^{-10},
\end{eqnarray}
where $n$ denotes the iteration step for solving the Poisson’s equation in Eq. (\ref{SMac-Poiss}). The first condition in Eq. (\ref{convergent-time-step}) is satisfied in the early iterations; therefore, the convergence condition is unnecessary. Note that only the convergent solution satisfies $\nabla\cdot{\vec V}\!=\!0$. Thus, the numerical solution of the LNS equation in Eq. (\ref{NS-eq-org}) is a steady-state solution, characterized by Eq. (\ref{SS-condition}).

\begin{figure}[ht]
\begin{center}
\includegraphics[width=10.5cm]{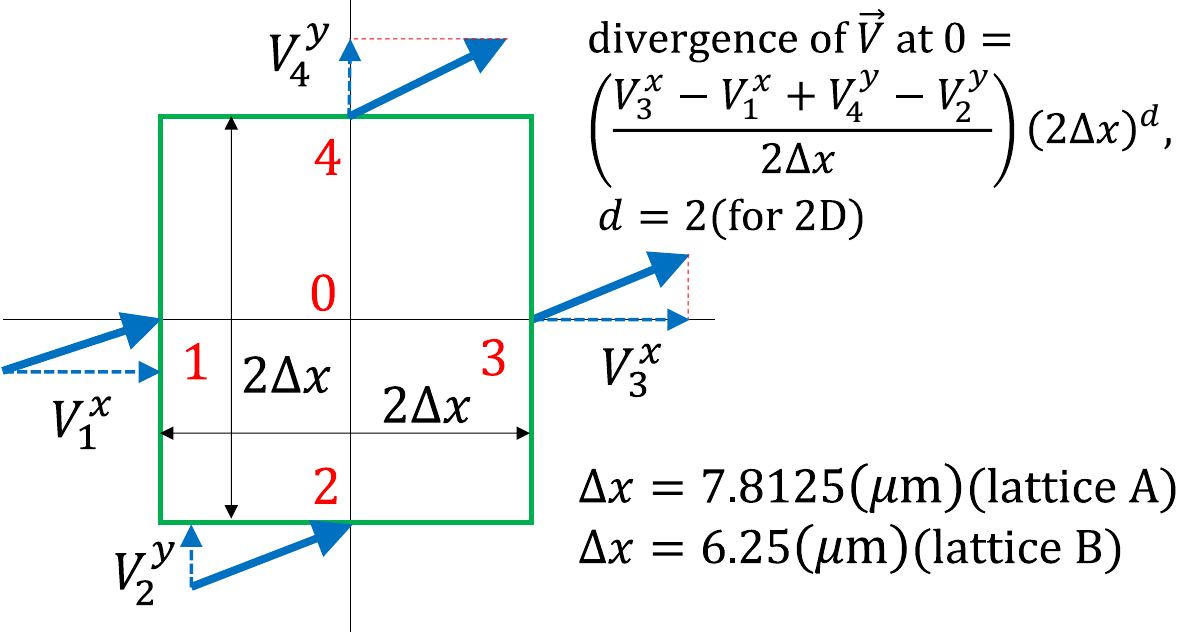}
\caption{ Illustration of divergence $(2{{\it \Delta} x})^2\nabla\cdot \vec{V}$ at lattice point 0, where the dimension is assumed to be $D\!=\!2$ for simplicity and where the lattice spacing ${\it \Delta} x$ is assumed in the 3D simulations on lattices A and in this paper. \label{fig-15} }
\end{center}
\end{figure}

To verify the accuracy of this technique, we calculate the lattice average ${\rm Div}_{\rm ab}$ of $|\nabla\cdot \vec{V}|$: ${\rm Div}_{\rm ab}\!=\!(2{\it \Delta} x)^3\sum_{ijk}|\nabla\cdot \vec{V}_{ijk}|/\sum_{ijk}1 ({\rm m^3/s})$, where $\sum_{ijk}1$ is the total number of internal lattice points and where $\nabla\cdot \vec{V}_{ijk}\!=\!(V_{i+1jk}^x\!-\!V_{i-1jk}^x\!+\!V_{ij+1k}^y\!-\!V_{ij-1k}^y\!+\!V_{ijk+1}^z\!-\!V_{ijk-1}^z)/(2{\it \Delta} x)$. Here, ${\rm Div}_{\rm ab}$
is considered to be the lattice average of the fluid volume flowing into or out of a cubic lattice enclosing a lattice point per second according to Gauss's theorem (see Fig. \ref{fig-15} for the 2D case).
For $D\!=\!0$, we numerically obtain $\sum_{ijk}|\nabla\cdot \vec{V}_{ijk}|\!=\!0 ({\rm 1/\beta s})$ for every time step $t$ (see Appendix \ref{append-B} for the simulation unit $\beta$); hence, ${\rm Div}_{\rm ab}\!=\!0$ for all lattice points, which implies that $\nabla\cdot \vec{V}\!=\!0$ on lattice B. We also have $\sum_{ijk}|\nabla\cdot \vec{V}_{ijk}|\!\simeq\! 7.9\times10^{-9}({\rm1/\beta s})\!=\! 1.9\times10^{-7}({\rm 1/s})$ on lattice A, which implies that the total divergence is given by $(2{\it \Delta} x)^3\sum_{ijk}|\nabla\cdot \vec{V}_{ijk}|\!\simeq\!7.4\times 10^{-4}({\rm \mu m^3/s})\!=\! 7.4\times 10^{-10}({\rm \mu g/s})$, where ${\it \Delta} x\!=\!7.8125\,{\rm (\mu m)}$, and $1({\rm \mu m^3})$ is replaced by $10^{-6}({\rm \mu g})$ because the density $\rho$ is considered to be the same as that of water $\rho\!=\!10^3({\rm kg/m^3})\!=\!10^{-3}({\rm g/mm^3})\!=\!10^{-12}({\rm g/\mu m^3})$. This value of the total divergence $7.4\times 10^{-10}({\rm \mu g/s})$, which implies ${\rm Div}_{\rm ab}\simeq 6.5\times 10^{-16}({\rm \mu g/s})$, is sufficiently small for the scales of protoplasmic streaming, where $\sum_{ijk}1\!=\!1,153,800$ is used for the total number of internal points of lattice A. Detailed information on the lattice geometry is presented in Section \ref{lattices}.

In the case of $D\!\not=\!0$, the total divergence fluctuates from one convergent configuration to another depending on the Brownian random forces. However, its mean value is almost independent of $D$, and its maximum value is approximately given by $\sum_{ijk}|\nabla\cdot \vec{V}_{ijk}|\!=\! 1\times 10^{-7}({\rm 1/\beta s})$, which is independent of lattices A and B; this value is comparable to the above-mentioned value for $D\!=\!0$ on lattice A. Thus, we find that the SMAC method is successful for the divergenceless condition when simulating the 3D LNS equation in Eq. (\ref{NS-eq-org}) under the condition of Eq. (\ref{SS-condition}).

\section{Physical units and simulation units\label{append-B}}
In the simulations, the physical units $({\rm m, s, kg})$ are changed to simulation units $({\rm \alpha m, \beta s, \lambda kg})$ using positive numbers $\alpha, \beta$ and $\lambda$. Using these numbers for $V_e, \nu_e,$ and $\rho_e$, we have the relations $V_e [{\rm m/s}]\!=\!V_e \beta/\alpha [{\rm \alpha m/(\beta s)}]$, $\nu_e [{\rm m^2/s}]\!=\!\nu_e \beta/\alpha^2 [{\rm (\alpha m)^2/(\beta s)}]$, and $\rho_e [{\rm kg/m^3}]\!=\!\rho_e \alpha^3/\lambda [{\rm \lambda kg/(\alpha m)^3}]$ in physical units. The right-hand sides of these relations can be written as $V_0 [{\rm \alpha m/(\beta s)}]$, $\nu_0 [{\rm (\alpha m)^2/(\beta s)}]$, and $\rho_0 [{\rm \lambda kg/(\alpha m)^3}]$ in the simulation units. Therefore,
\begin{eqnarray}
	\label{numbers-for-units}
	\alpha=\frac{\nu_e}{\nu_0}\frac{V_0}{V_e}, \quad \beta=\frac{\nu_e}{\nu_0}\left(\frac{V_0}{V_e}\right)^2, \quad \lambda=\frac{\rho_e}{\rho_0}\left(\frac{\nu_e}{\nu_0}\frac{V_0}{V_e}\right)^3.
\end{eqnarray}

In addition to these numbers, we need positive numbers $\gamma$ and $\delta$ for the lattice and time discretization, such that $n_X\to \gamma n_X$ and $n_T\to \delta n_T$ for the physical quantities to be independent of $n_X$ and $n_T$. In this expression, $n_X$ and $n_T$ are connected to the lattice spacing ${\it \Delta}x ({\rm m})$ and discrete time step ${\it \Delta}t ({\rm s})$, respectively, such that ${\it \Delta}x ({\rm m})\!=\!d_e/n_X$ and ${\it \Delta}t ({\rm s})\!=\!\tau_e/n_T$, where $\tau_e$ is the relaxation time \cite{Coffey-Kalmykov-CP1993,Feldmanetal-Wiley2006}. In this paper, we do not provide details of this problem for $n_X$ and $n_T$, and $\gamma$ and $\delta$ are fixed to $\gamma\!=\!1$ and $\delta\!=\!1$. This problem is studied in Ref. \cite{Noro-etal-2021}, where the ${\it \Delta}t$ dependence is studied under $\gamma\!=\!1$.

We assume the values given in Table \ref{table-B1} for the unit change.
\begin{table}[htb]
\caption{Values for the change in physical units and simulation units. \label{table-B1}}
\begin{center}
 \begin{tabular}{cccccccccccc }
 \hline
$\alpha$ && $\beta$ && $\lambda$ && $\gamma$ && $\delta$ \\ \hline
 $2 \times10^{-6}$ && $4\times 10^{-2}$ && $8\times10^{-12}$ && $1$ && 1 \\ \hline
\end{tabular}
\end{center}
\end{table}
Using the assumed numbers $\alpha$, $\beta$, and $\gamma$ in Table \ref{table-B1}, we have
$V_0\!=\!V_e \beta/\alpha\!=\!(50\!\times\!10^{-6})(4 \!\times\!10^{-2})/(2\!\times\!10^{-6})\!=\!1 [{\rm \alpha m/(\beta s)}]$, $\nu_0\!=\!\nu_e \beta/\alpha^2\!=\!(1\!\times\!10^{-4})(4 \!\times\!10^{-2})/(2\!\times\!10^{-6})^2 \!=\!1\!\times\!10^{-6}[{\rm (\alpha m)^2/(\beta s)}]$, and $\rho_0\!=\!\rho_e \alpha^3/\lambda\!=\!(1 \!\times\!10^{3})(2\!\times\!10^{-6})^3/(8\!\times\!10^{-12})\!=\! 1 \!\times\!10^{-3}[{\rm \lambda kg/(\alpha m)^3}]$.

The strength of the random force $D_e({\rm m^2/s^3})$ is expressed by $D ({\rm (\alpha m)^2/(\beta s)^3})$ in the simulation units such that
\begin{eqnarray}
\label{D-units-change}
\begin{split}
&D_e({\rm m^2/s^3})=\alpha^{-2}\beta^3 D_e ({\rm (\alpha m)^2/(\beta s)^3})=1.6\times 10^5 D_e({\rm (\alpha m)^2/(\beta s)^3}),\\
\Leftrightarrow & D=\alpha^{-2}\beta^3 D_e=1.6\times 10^5 D_e({\rm (\alpha m)^2/(\beta s)^3}) \\
 \Leftrightarrow & D_e=6.25\times 10^{-6} D ({\rm m^2/s^3}).
\end{split}
\end{eqnarray}
In the simulations, $D(=\!1.6\times 10^5 D_e)({\rm (\alpha m)^2/(\beta s)^3})$ is varied implying that $D_e({\rm m^2/s^3})$ is varied. 

The lattice spacing ${\it \Delta}x_0$ in the simulation units is given by ${\it \Delta}x_0\!=\!\alpha^{-1}\frac{d_e}{n_X}\!=\!(2\!\times\!10^{-6})^{-1}(500\!\times\!10^{-6})/(2R)$, where the diameter $2R$ of the cylinder is assumed to be $n_X$, which is the total number of discretizations introduced for a regular square lattice of size $L\!\times\! L$ with $L\!=\!n_X{\it \Delta}x_0$. For $n_X\!=\!2R\!=\!64$ on lattice A ($n_X\!=\!2R\!=\!80$ on lattice B), we have ${\it \Delta}x_0\!=\!3.90625$ (${\it \Delta}x_0\!=\!3.125$)  in the simulation unit ${\rm (\alpha m)}$ and ${\it \Delta}x\!=\!7.8125$ (${\it \Delta}x\!=\!6.25$) in the physical unit $ ({\rm \mu m})$.

The discrete time step ${\it \Delta}t_0$ can also be expressed by ${\it \Delta}t_0\!=\!\beta^{-1}\frac{\tau_e}{n_T}$ using the macroscopic relaxation time $\tau_e$ and total number of time discretizations $n_T$. However, $\tau_e$ is not always given; therefore, we assume that ${\it \Delta}t_0\!=\!5\!\times\!10^{-7}({\rm \beta s})$  in the simulation unit and ${\it \Delta}t\!=\!2\!\times\!10^{-8}({\rm s})$ in the physical unit.

\section{Calculation technique for the velocity distribution \label{append-C}}
\begin{figure}[h!]
\begin{center}
\includegraphics[width=9.5cm]{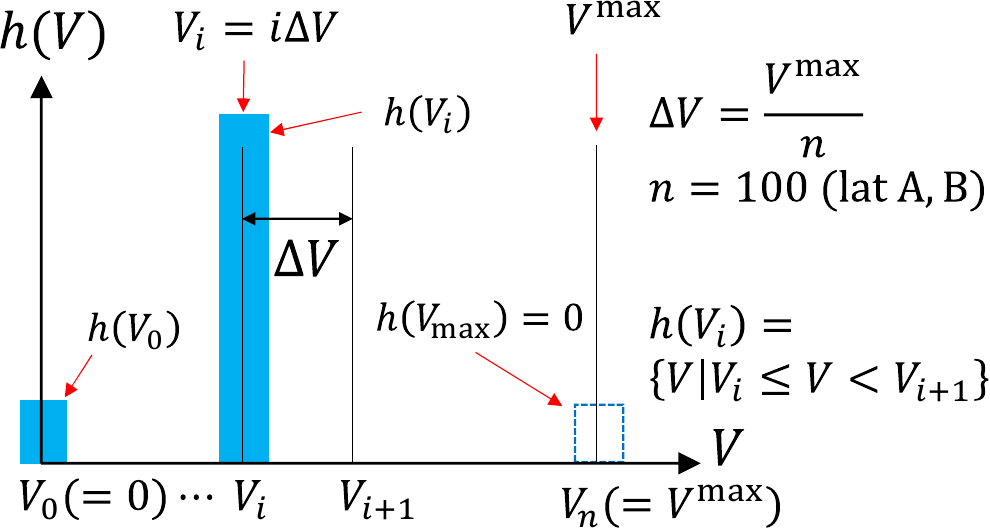}
\caption{The histogram $h(V)$ of velocity $V$ numerically calculated for all cross-sections of lattices A and B in Figs. \ref{fig-3}(a), (b). A small ${\it \Delta}V$ is fixed to ${\it \Delta}V\!=\!V^{\rm max}/n$ using the maximum velocity $V^{\rm max}$ and the number $n$, which is $n\!=\!100$ for both lattices A and B for nonzero $D$. For $D\!=\!0$, smaller $n$ values are assumed. The histogram $h(V_i)$ is obtained by counting the total number of lattice points at which the velocity $V$ satisfies the condition $ V_i\!\leq\!V\!<\!V_{i+1}$. The normalization of $V$ is defined as $V\!\to\! V/V^{\rm max}$, where $V^{\rm max}$ depends on $D$. The histogram $h(V_z)$ is obtained using the same procedure as for $h(V)$ by replacing $V\!\to\!|V_z|$. Therefore, the horizontal axis is denoted as $|V_z|$ for the plots of $h(V_z)$.
 \label{fig-16} }
\end{center}
\end{figure}

In this Appendix, we present the calculation technique for the velocity distribution $h(V)$ in detail. The same technique is applied to $h(V_z)$ by replacing $V\!\to\!|V_z|$. The total number of cross-sections is 360 in lattice A and 80 in lattice B, and the total number of sample configurations is 100 for lattices A and B, which are smaller than 1000 for the calculation of $V_z(r,\theta)$ at three different $\theta$. However, $h(V)$ and $h(V_z)$ are calculated for all cross-sections, and hence, the statistics is relatively high compared to the case of $V_z(r,\theta)$, which is calculated only on the cross section at $z\!=\!L/2$. W should note that the velocities of the boundary points are not included in the sample configurations.

To calculate the distribution $h(V)$, we first search for the maximum velocity $V^{\rm max}$ in all sample configurations for each value of $D$ and define the width of the velocity by ${\it \Delta}V\!=\!V^{\rm max}/n$, where $n\!=\!100$ for both lattices A and B (Fig. \ref{fig-16}). The histogram $h(V_i), (0\!\leq\!i\!\leq\!n)$ at $V_i\!=\!i {\it \Delta}V$ is calculated by counting the number of lattice sites where the velocity $V$ satisfies $V_i\!\leq\!V\!<\!V_{i+1}$. The plot of $h(V)$ for the normalized velocity is obtained by replacing $V$ with $V\!\to\!V/V^{\rm max}$.

\section{Impulse action of the Brownian force and fluctuation dissipation relations \label{append-D}}
First, to describe the action of a random force on fluid velocity, we introduce
\begin{eqnarray}
\label{Integral-GRF}
\vec {H}_{i}(t; {\it \Delta} \tau)=\int_{t}^{t+{\it \Delta}\tau}\vec \eta_{i}(t) dt,
\end{eqnarray}
where the suffix $i$ denotes a 3D lattice site. $\vec{H}_{i}(t; {\it \Delta} \tau)$ plays a role in an impulse action, which is ${\it \Delta} \tau$ times the mean value of $ {\vec \eta}$ in Eq. (\ref{NS-eq-org}) from $t$ to $t\!+\!{\it \Delta} \tau$ at $t$ \cite{Egorov-etal-POF2020}, satisfying the relations
\begin{eqnarray}
\label{Brownian-motion}
\left\langle {\vec H}_{i}(t; {\it \Delta} \tau)\right\rangle=0, \quad \left\langle H_{i}^{a \,2}(t; {\it \Delta} \tau)\right\rangle\!=\!2D_\tau{\it \Delta} \tau, \;(a=1,2,3),
\end{eqnarray}
where $D_\tau$ corresponds to $D$ in Eq.  (\ref{Gaussian-random-force}), and $\langle * \rangle$ denotes the mean value of many samples or ensemble average and is the same as that in Eq. (\ref{Gaussian-random-force}). Assuming the integrand $\vec \eta_{i}(t)$ in Eq. (\ref{Integral-GRF}) to be constant and finite during ${\it \Delta} \tau$, we rewrite the right-hand side of Eq. (\ref{Integral-GRF}) as $\vec{H}_{i}(t; {\it \Delta} \tau)\!=\!\vec \eta_{i}(t){\it \Delta} \tau$, where the same symbol $\vec \eta_{i}(t)$ is used. 
 Using the second part of Eq. (\ref{Brownian-motion}), we obtain the absolute value of the Brownian force $\left| \eta_{i}^a(t)\right|\!=\!\sqrt{2D_\tau/{\it \Delta} \tau}$ in the sense of squared mean value.

 If the time step ${\it \Delta} t$ is used instead of ${\it \Delta} \tau$ in Eq. (\ref{Integral-GRF}), the relations in Eq. (\ref{Brownian-motion}) are written as 
  \begin{eqnarray}
 	\label{Brownian-motion-dt}
 	\left\langle {\vec H}_{i}(t; {\it \Delta} t)\right\rangle=0, \quad \left\langle H_{i}^{a \,2}(t; {\it \Delta} t)\right\rangle\!=\!2D{\it \Delta} t, \;(a=1,2,3),
 \end{eqnarray}
where we write $D_\tau$ as $D$, and we have the expression 
 \begin{eqnarray}
 	\label{eta-component}
 	\left|\eta_{i}^a(t)\right|=\sqrt{2D/{\it \Delta} t}
 \end{eqnarray}
for each component of the  Brownian force. $D$  is the strength of one impulse action, numerically introduced by $\vec{\eta}_{i}{\it \Delta} t\!=\!\sqrt{2D{\it \Delta} t}\vec g_{i}(t)$ for $\vec{V}_i$ fluctuations, where $\vec g_{i}(t)\!=\!(g^1_{i},g^2_{i},g^3_{i})$ is the Gaussian random number of mean 0 and variance 1 satisfying $\langle g_i^a\rangle\!=\!0$, $\langle g_i^{a\,2}\rangle\!=\!1$, $(a\!=\!1,2,3)$.  As described in Appendix \ref{append-A},  ${\it \Delta} t$ is introduced to obtain a convergent $\vec{V}_i$ on which impulse $\vec{\eta}_{i}{\it \Delta} t$ is applied at $t$, and  the same impulse is applied at every ${\it \Delta} t$ during the iterations until the convergent configuration of $\vec{V}_i$ is obtained. Therefore, the impulse action  $\vec{H}_{i}(t; {\it \Delta} \tau)$, where ${\it \Delta} \tau\!=\!n{\it \Delta} t$, $(n=1,2,\cdots)$, is proportional to ${\it \Delta} \tau$:  For ${\it \Delta} \tau\!=\!2{\it \Delta} t$, $\vec{H}_{i}(t; {\it \Delta} \tau)\!=\!\int_t^{t+2{\it \Delta} t}\vec \eta_{i}(t) dt\!=\!\int_t^{t+{\it \Delta} t}\vec \eta_{i}(t) dt\!+\!\int_{t+{\it \Delta} t}^{t+2{\it \Delta} t}\vec \eta_{i}(t\!+\!{\it \Delta} t) dt\!=\!2\vec{H}_{i}(t; {\it \Delta} t)$. Consequently, $D_{\tau}$ is proportional to  ${\it \Delta}t$ because the second of Eq. (\ref{Brownian-motion}):
\begin{eqnarray}
\label{D-correspondence}
D_{\tau}=D \frac{{\it \Delta} \tau}{{\it \Delta} t}.
\end{eqnarray}
In this paper, we use $D$ for the strength of the numerically introduced random force in the text  for simplicity to save symbols.

If ${\it \Delta} \tau$ is used to define $\vec{H}_{i}(t; {\it \Delta} \tau)$ in Eq. (\ref{Integral-GRF}), the Brownian force can be modified from $\left| \eta_{i}^a(t)\right|_{{\it \Delta} t}\!=\!\sqrt{2D/{\it \Delta} t}$ to $\left| \eta_{i}^a(t)\right|_{{\it \Delta} \tau}\!=\!\sqrt{2D_\tau/{\it \Delta} \tau}$. We can easily verify that $\left| \eta_{i}^a(t)\right|_{{\it \Delta} t}\!=\!\left| \eta_{i}^a(t)\right|_{{\it \Delta} \tau}$ by using Eq. (\ref{D-correspondence}). Therefore, the Brownian impulse in the discrete LNS equation for time step ${\it \Delta} t$  is independent of ${\it \Delta} \tau$: $\vec \eta_{i}(t)|_{{\it \Delta} t}{\it \Delta} t\!=\!\vec \eta_{i}(t)|_{{\it \Delta} \tau}{\it \Delta} t$. This implies that the effects of Brownian force $D_\tau$ during ${\it \Delta} \tau(>\!{\it \Delta} t)$ can be simulated by using $D$ and ${\it \Delta} t$. Therefore, ${\it \Delta} \tau$ can be arbitrarily chosen under conditions ${\it \Delta} \tau\!>\!{\it \Delta} t$ and Eq. (\ref{D-correspondence}).

Because ${\it \Delta} \tau$ is independent of  ${\it \Delta} t$,  it is possible to regard ${\it \Delta} \tau$ as a relaxation time to evaluate the net effect of Brownian forces corresponding to  ${\it \Delta} \tau$  in a specific system in the framework of our modeling scheme using $D$ corresponding to ${\it \Delta}t$. Therefore, the formula in Eq. (\ref{D-correspondence}) is used to evaluate the strength $ D_\tau$ of the Brownian force for $\vec{V}$ of the protoplasmic streaming with relaxation time ${\it \Delta} \tau$ from a physically relevant $D$ estimated by comparing the simulated  and reported experimental velocity distributions. 

Now, we approximately evaluate $D_{\tau}$ of the protoplasmic streaming from the discrete form of the LNS equation of Landau and Lifschitz in Ref. \cite{Landau-Lifschitz-StatPhys}, where $\delta(\vec{r}\!-\!\vec{r}^{\,\prime})$ is replaced by $\delta_{ij}/{\rm v}_i$ with a small volume ${\rm v}_i$. Note that this small volume limit implies that  $\delta(\vec{r}\!-\!\vec{r}^{\,\prime})$ can be replaced by $\delta_{ij}/{\rm v}_i$ if we use lattices with sufficiently small lattice spacing ${\it \Delta} x$. This replacement $\delta(\vec{r}\!-\!\vec{r}^{\,\prime})\!\to\!\frac{\delta_{ij}}{({\it \Delta} x)^3} ({\it \Delta} x\!\to\! 0)$  is suitable for the numerical simulations of the LNS equation. Moreover, $D$ depends on ${\it \Delta} x$ as described in the text;  hence, when ${\it \Delta} x$ and $D$ are fixed,  ${\it \Delta} x$ can be varied if $D$ is suitably varied such that the results remain unchanged. For these reasons, we first fix ${\it \Delta} x$ and then vary ${\it \Delta} x$-dependent $D$ to find a suitable $D$ for the experimentally observed velocity distribution. This experimentally relevant $D$ can be compared with $D_\tau$  expected from the fluctuation dissipation relation using a suitable ${\it \Delta} \tau$ and Eq. (\ref{D-correspondence}).

In the LNS equation for velocity $\vec{V}$ in Ref. \cite{Landau-Lifschitz-StatPhys}, a random force term is  given by $\rho^{-1}\frac{\partial s_{ab}}{\partial x_b}(=\!\rho^{-1}\sum_{b=1}^3\frac{\partial s_{ab}}{\partial x_b})$, which corresponds to $\eta^a(t)$ in our LNS equation in Eq. (\ref{NS-eq-org}).
Using  the expressions 
\begin{eqnarray}
\begin{split}
&s_{ab}(t,\vec{r}+{\it\Delta}\vec{x}_b)=s_{ab}(t,\vec{r})+\frac{\partial s_{ab}}{\partial x_b}(t,\vec{r}){\it\Delta}x,\;(a,b=1,2,3),\\
&{\it\Delta}\vec{x}_1=({\it\Delta}x,0,0),\; {\it\Delta}\vec{x}_2=(0,{\it\Delta}x,0),\;{\it\Delta}\vec{x}_3=(0,0,{\it\Delta}x),
\end{split}
\end{eqnarray}
we find 
\begin{eqnarray}
\begin{split}
\label{FD-A}
({\it\Delta}x)^{2}\left\langle \frac{\partial s_{ab}}{\partial x_b}(t,\vec{r}) \frac{\partial s_{ab}}{\partial x_b}(t^\prime,\vec{r})\right\rangle
=&\left\langle \sum_{b}s_{ab} (t,\vec{r}+{\it\Delta}\vec{x}_b)\sum_{b}s_{ab}(t^\prime,\vec{r}+{\it\Delta}\vec{x}_b)\right\rangle\\
&+\left\langle \sum_{b}s_{ab} (t,\vec{r})\sum_{b}s_{ab}(t^\prime,\vec{r})\right\rangle.
\end{split}
\end{eqnarray}
The left-hand side  of Eq. (\ref{FD-A}) corresponds to $({\it\Delta}x)^{2}\rho^2\langle \eta^a(t,\vec{r})\eta^b(t^\prime,\vec{r})\rangle$ from Eq. (\ref{LL-corespondence}). Since the magnitude of ${\it\Delta}x\frac{\partial {s}_{ab}}{\partial x_b}(t,\vec{r})$ is estimated to be ${\it\Delta}x|\frac{\partial {s}_{ab}}{\partial x_b}(t,\vec{r})|\!=\!{\it\Delta}x\rho|\eta^a(t,\vec{r})| \!=\!{\it\Delta}x\rho \sqrt{2D{\it \Delta} t}$ using Eq. (\ref{eta-component}), the left-hand side of    Eq. (\ref{FD-A}) is given by $({\it\Delta}x)^2\rho^2 2D_\tau{\it \Delta} \tau$, where  Eq. (\ref{D-correspondence}) is used.
The first and second terms on the right-hand side are directly calculated by the fluctuation dissipation formula in Ref. \cite{Landau-Lifschitz-StatPhys} for incompressible and viscous fluids 
with viscosity $\rho\nu$
\begin{eqnarray}
\left\langle s_{ab} (t,\vec{r})s_{cd}(t^\prime,\vec{r})\right\rangle=\frac{2k_BT\rho\nu}{{\rm v}_i}\left(\delta_{ac}\delta_{bd}+\delta_{ad}\delta_{bc}\right)\delta(t-t^\prime),
\end{eqnarray}
where $\delta(\vec{r}\!-\!\vec{r}^{\,\prime})$ is replaced by $1/{\rm v}_i$ at $\vec{r}$.
We find that the first term of the right-hand side of Eq. (\ref{FD-A}) is 
$\langle \sum_{b}s_{ab} (t,\vec{r}\!+\!{\it\Delta}\vec{x}_b)\sum_{b}s_{ab}(t^\prime,\vec{r}\!+\!{\it\Delta}\vec{x}_b)\rangle\!=\!\frac{8k_BT\rho\nu}{{\rm v}_i{\it\Delta}\tau}$, and the second is $\langle \sum_{b}s_{ab} (t,\vec{r})\sum_{b}s_{ab}(t^\prime,\vec{r})\rangle\!=\!\frac{8k_BT\rho\nu}{{\rm v}_i{\it\Delta}\tau}$, 
 and therefore, the right-hand side of    Eq. (\ref{FD-A}) is  $\frac{16k_BT\rho\nu}{{\rm v}_i{\it\Delta}\tau}$. Note that $\delta(t\!-\!t^\prime)$ is replaced by $1/{\it\Delta}\tau$ because ${\it\Delta}t$ should be replaced by ${\it\Delta}\tau$ when our modeling scheme is applied to a physical system as described above. Thus, we obtain $\rho^2 2D_{\tau}/{\it \Delta} \tau\!=\!{\frac{16k_BT\rho\nu}{{\rm v}_i{\it\Delta}\tau}}/{\it\Delta}x$.   Replacing ${\rm v}_i\!\to\!({\it\Delta}x)^3$, we have 
\begin{eqnarray}
\label{D-vs-nu}
D_\tau = \frac{8\nu k_BT}{\rho({\it\Delta}x)^5}.
\end{eqnarray}
The value is   $D_\tau\!=\!1.14\!\times\! 10^{-1}\!\simeq\!0.1({\rm m^2/s^3})$ for $\rho_e\!=\!1\!\times\!10^3 ({\rm kg/m^3})$,  $\nu_e\!=\!1\!\times\!10^{-4} ({\rm m^2/s})$,  ${\it\Delta}x\!=\!7.8125\!\times\!10^{-6}({\rm m})$ for lattice A in Table \ref{table-2}, and $k_BT\!=\!4.14\!\times\!10^{-21} (\rm{m^2kg/s^{2}})$ for room temperature $T\!=\!300 ({\rm K})$. 
The symbols with the subscript $e$ denote the experimentally observed or observable quantities and the quantities in the physical unit. This value $D_{\tau}\!\simeq\!0.1({\rm m^2/s^3})$ is comparable with $D_{e,\tau}\!=\!0.08({\rm m^2/s^3})$ for $D\!=\!50$ and $D_{e,\tau}\!=\!0.16({\rm m^2/s^3})$ for $D\!=\!100$  in Table \ref{table-4}, where $D_{e,\tau}$ is obtained by the formula $D_{e,\tau}\!=\!D_{e}\frac{{\it \Delta} \tau}{{\it \Delta} t}$  in Eq. (\ref{D-correspondence}). In this formula, ${\it \Delta} t\!=\!2\!\times\!10^{-8}({\rm s})$ is used, and we assume ${\it \Delta} \tau\!=\!5\!\times\!10^{-4}({\rm s})$.

An analog of the relaxation time ${\it \Delta} \tau_{\rm sim}$ is numerically obtained by ${\it \Delta} \tau_{\rm sim}\!=\!\langle n_{\rm itr}\rangle {\it \Delta t}$, where $\langle n_{\rm itr}\rangle$ is the mean value of the total number of iterations for convergence with the initial configuration of $\vec{V}_i\!=\!(0,0,0)$ for all positions $i$. ${\it \Delta} \tau_{\rm sim}$ is approximately equal to ${\it \Delta} \tau_{\rm sim}\!=\!2.035\!\times\!10^{-2}({\rm \beta s})\!\simeq\!8.1\!\times\!10^{-4}({\rm s})$ for $D\!=\!50$ and ${\it \Delta} \tau_{\rm sim}\!=\!2.136\!\times\!10^{-2}({\rm \beta s})\!\simeq\!8.5\!\times\!10^{-4}({\rm s})$ for $D\!=\!100$.  The values of ${\it \Delta} \tau_{\rm sim}$ are almost independent of the initial configuration of $\vec{V}_i$. Let $\{\vec{V}_i(t_n),(n\!=\!1,2,\cdots)\}$ represent a series of convergent configurations. If the initial configuration $\vec{V}_i$ at $t_{n+1}$ is fixed to the convergent configuration $\vec{V}_i(t_n)$ instead of $\vec{V}_i\!=\!(0,0,0)$, ${\it \Delta} \tau_{\rm sim}$ remains almost unchanged. Because the assumed relaxation time, ${\it \Delta} \tau\!=\!5\!\times\!10^{-4}({\rm s})$ is the equilibration time of fluid volume $({\it \Delta} x)^3$ for a random Brownian force, we consider  ${\it \Delta} \tau\!=\!5\!\times\!10^{-4}({\rm s})$ to be reasonable because it is comparable to  ${\it \Delta} \tau_{\rm sim}\!\simeq\!8\!\times\!10^{-4}({\rm s})$.  

Another expression for the fluctuation dissipation relation similar to Eq. (\ref{D-vs-nu}) can also be obtained by considering fluid volume $({\it \Delta}x)^3$ as a particle of mass $m\!=\!\rho ({\it \Delta}x)^3$ and of size ${\it \Delta}x$. Let us assume that this particle of mass $m$ flows in the fluid with a random Brownian force $\vec{R}_i(t)$ acting on the particle at position $i$. It is natural to assume the fluctuation dissipation relation for $\vec{R}_i(t)$:
\begin{eqnarray}
\label{Langevin-force}
\left\langle R_i^a(t)R_j^b(t^\prime)\right\rangle = 2\gamma k_BT\delta_{ij}\delta^{ab}\delta(t-t^\prime), \quad 
\end{eqnarray}
where $\gamma$ denotes a friction constant between the particle and fluids \cite{Nagahiro-etal-PRE2007}. This random force $\vec{R}_i(t)$ has the unit of force. Therefore, we have a relation $m^{-1}\vec{R}_i(t)\!=\!\vec{\eta}_i(t)$ from the correspondence between the Langevin equation for the fluid particle and the LNS equation for the velocity in this paper. Therefore, comparing  the relation in Eq. (\ref{Langevin-force}) for $\vec{R}_i(t)$ and that in Eq. (\ref{Gaussian-random-force}) for  $\vec{\eta}_i(t)$, we have 
\begin{eqnarray}
\label{correspondence-Langevin}
m^{-2}2\gamma k_BT= 2D_{\tau}. 
\end{eqnarray}
The friction constant $\gamma$ is replaced by the kinematic viscosity $\nu$ using the Stokes formula  $\gamma\!=\!6\pi \rho \nu {\it \Delta}x$. Thus, we obtain
\begin{eqnarray}
\label{D-vs-nu-Langev}
D_{\tau} = \frac{6\pi\nu k_BT}{\rho({\it\Delta}x)^5}\simeq \frac{19\nu k_BT}{\rho({\it\Delta}x)^5},
\end{eqnarray}
which is approximately twice as large as that of Eq. (\ref{D-vs-nu}).

 \section{Polynomial function fitting the velocity distribution at $D\!=\!0$ \label{append-E}}
 The polynomial function $h_0(u)$ of Eq. (\ref{theoretical}) corresponding to $h(V_z)$ at $D\!=\!0$ of lattice A in Fig. \ref{fig-6}(a) is given by
 \begin{eqnarray}
 	\label{polynomial}
 	\begin{split}
 		& h_0(u)=\left\{ \begin{array}{@{\,}ll}
 			\sum_{i=0}^8 a_i u^i, \quad\quad (1\geq u\geq 0) \\
 			\sum_{i=0}^8 a_i (-u)^i \quad (-1\leq u<0)
 		\end{array} 
 		\right., \\
 		&a_0=3.25\times 10^{-1}, \;a_1=1.25\times 10^{-1}, \;a_2=-1.78, \\
 		&a_3=1.49\times 10^{1}, \;a_4=-5.58\times 10^{1}, \;a_5=1.18\times 10^{2}, \\
 		&a_6=-1.40\times 10^{2}, \;a_7=8.79\times 10^{1}, \;a_8=-2.24\times 10^{1}.
 	\end{split}
 \end{eqnarray}
 
\section{Convergence of the time evolution iteration   \label{append-F}}
\begin{figure}[ht]
	\begin{center}
		\includegraphics[width=11.5cm]{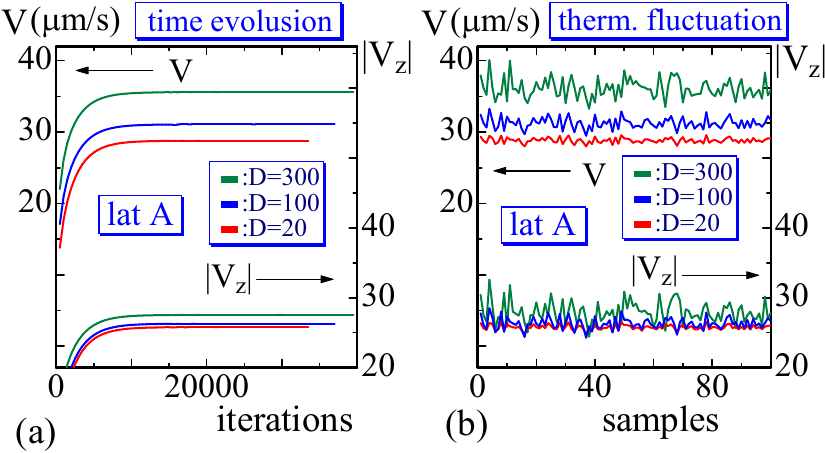}
		\caption{(a) Discrete time evolution of lattice averages of $V$ and $|V_z|$ calculated at all internal vertices for $D\!=\!300$, 	$D\!=\!100$ and	$D\!=\!20$, (b)	dependence of the convergent $V$ and $|V_z|$ on samples representing the thermal fluctuations. \label{fig-17} }
		
	\end{center}
\end{figure}

In this Appendix, we examine whether the total momentum remains unchanged during  the time-evolution process. As discussed in Section \ref{FD-ReL}, the Brownian force $\vec{\eta}_i$ in Eq. (\ref{NS-eq-org}) is expected to play a role in the fluid dynamic interaction in the sense that  $\vec{\eta}_i$ is not contradictory to global momentum conservation. The lattice averages $V$ and $|V_z|$ versus the number of iterations are plotted in Fig. \ref{fig-17}(a), where $V\!=\!\sum_{i}\|\vec{V}_{i}\|/\sum_{i}1$ and $|V_z|\!=\!\sum_{i}|V^z_{i}|/\sum_{i}1$ are calculated at all the internal lattice sites $i$. As described in Section \ref{LNS-equation}, the discrete time step does not always correspond to real-time evolution, because the divergence-less condition is not always satisfied. However, the plots show that $V$ and $|V_z|$ remain unchanged soon after the start of the simulations, implying that the total momentum is conserved. 

The convergent values depend on the Brownian force $\vec{\eta}$ and are expected to fluctuate. To observe these fluctuations, we plot 100 samples of $V$ and $|V_z|$ in Fig. \ref{fig-17}(b), where the data are connected by solid lines for a clear visualization. The fluctuations correspond to thermal fluctuations, and their mean values are meaningful physical quantities in statistical mechanical analysis. Thus, we confirm that the thermally fluctuating total momentum is conserved in terms of the mean value even though the random force  $\vec{\eta}_i$ appears as an external force. This observation indicates that the correspondence between $\vec{\eta}_i$ and the fluid dynamic force of Landau-Lifshitz in Eq. (\ref{LL-corespondence}) is well-defined.

\section*{References}

\end{document}